%% file: WSarXivFeb25.tex
\documentclass{jfm1}
\usepackage{graphicx}
\usepackage{natbib}
\usepackage{amssymb,amsmath}
\usepackage{color}
\input colors.defs

\input kerrcommands.tex

\newcommand{\bpurp}[1]{#1} 
\newcommand{\bGree}[1]{} 
\newcommand{\Lap}{\mbox{\boldmath$\triangle$}}
\newcommand{\snm}[3]{\|{#1}\|_{#2}^{#3}} 
 
\newcommand\etall{\mbox{\textit{et al.}}}

\newcommand{\emailnotes}[1]{}
\newcommand{\biband}{\&~}
\newcommand{\authone}[2]{#2,~#1}
\newcommand{\authtwo}[4]{#2,~#1,~\&~#4,~#3}
\newcommand{\auththr}[6]{#2,~#1,~#4,~#3,~\&~#6,~#5}
\newcommand{\authfour}[8]{#2,~#1,~#4,~#3,~#6,~#5,~\&~#8,~#7} 
\newcommand{\authmanytwo}[4]{#2,~#1,~#4,~#3,} 
\newcommand{\authmanythr}[6]{#2,~#1,~#4,~#3,~#6,~#5,}

\newcommand{\yjour}[6]{ #1~ #6. {\em #2} {\bf #3}, #4#5.}

\newcommand{\ybook}[3]{ #1~ {\em #2}. #3.}

\title[Navier-Stokes bounds and scaling for compact trefoils]{Navier-Stokes bounds and scaling for compact trefoils in large $(2\ell\pi)^3$ domains.}

\author[R. M. Kerr]
{Robert M. Kerr
  \thanks{Email address for correspondence: Robert.Kerr@warwick.ac.uk},\ns}

\affiliation{Department of Mathematics, University of Warwick,
Coventry CV4 7AL, United Kingdom}

\pubyear{}
\volume{}
\pagerange{}
\begin{document}

\maketitle

\begin{abstract} 
For a perturbed trefoil vortex knot evolving under the Navier-Stokes equations, a sequence of $\nu$-independent times $t_m$ are identified corresponding to a set of scaled, volume-integrated vorticity moments $\nu^{1/4}{\cal O}_{V1}$ with this hierarchy $t_\infty\le\dots\le t_m\dots t_1=t_x\approx40$ and ${\cal O}_{Vm}=(\int_{V\ell}|\omega|^{2m}dV)^{1/2m}$. For $Z(t)={\cal O}_{V1}(t)\bigr)^2$ the volume-integrated enstrophy, convergence of $\sqrt{\nu}Z(t)$ at $t_x=t_1$ marks the end of the reconnection scaling phase. Physically, reconnection follows from the formation of a double vortex sheet, then a knot, which splits into spirals. $Z$ then accelerates, leading to approximate finite-time $\nu$-independent convergence of the energy dissipation rate $\epsilon(t)=\nu Z(t)$ at $t_\epsilon\sim 2t_x$ and sustained over a finite span  $\Delta T_\epsilon\searrow 0.5 t_\epsilon$, giving Reynolds number independent finite-time,
dissipation, $\Delta E_\epsilon=\int_{\Delta T_\epsilon}\epsilon dt$, and thus satisfying one definition for a {\it dissipation anomaly}. Evidence for transient Kolmogorov-like enstrophy spectra is found over ${\Delta T_\epsilon}$. A critical factor in achieving these temporal convergence laws is how the domain $V_\ell=(2\ell\pi)^3$ is increased as $\ell\sim\nu^{-1/4}$, for $\ell=2$ to 6, then to $\ell=12$, as $\nu$ decreases.  $(2\ell\pi)^3$ domain compatibility with established $(2\pi)^3$ mathematics in appendix allows small $\nu$ Navier-Stokes solutions. Two spans of $\nu$ are considered. Over the first factor of 25 decrease in $\nu$, all of the $\nu^{1/4}{\cal O}_{Vm}(t)$ converge to their respective $t_m$. For the next factor of 5 decrease in $\nu$, $\ell$ is increased to $\ell=12$, there is only convergence of $\nu^{1/4}\Omega_{V\infty}(t)$ to $t_\infty$ and  later $\sqrt{\nu}Z(t)$ convergence at $t_1=t_x$ and $\epsilon(t)$ over $t\sim t_\epsilon$.
\end{abstract}
\vspace{-8mm}
\section{Background \label{sec:back}}

Finite-energy dissipation is ubiquitous in turbulent, large Reynolds number, 
geophysical and engineering flows. Meaning that in these systems, observations show
that their turbulent flows always develop Reynolds number, or viscosity $\nu$, 
independent dissipation \citep{Sreeni1984,Vassilicos2015}. Forced simulations can generate
finite-dissipation, but it is unclear what
the pre-cursors to finite-dissipation might be if the flow is unforced. 
Can trefoil reconnection calculations \citep{KerrJFMR2018,KerrPRF2023} fill that gap?

To fill that gap, this paper introduces this pre-cursor for trefoil knot analysis: 
Temporal convergence of $\nu^{1/4}$ scaled, volume-integrated vorticity moments,
$\nu^{1/4}{\cal O}_{Vm}(t)$, with
\vspace{-2mm}
\EQL{eq:OmegamV} {\cal O}_{Vm}(t)=\left(\int_{\cal V}|\omega|^{2m} dV\right)^{1/2m}\,\text{and}
\quad {\cal O}_{V1} \leq\dots\leq d_m {\cal O}_{Vm}(t)\leq\dots\leq
d_\infty{\cal O}_{\infty}(t)\,. \EN
\vspace{-3mm}

\noindent
The hierarchy of convergence times is $t_\infty\le\dots\le t_m\dots t_1=t_x$ and
precedes the post-reconnection $t>t_x$ accelerated growth of the volume-integrated
enstrophy $Z(t)={\cal O}^2_{V1}(t)$ \eqref{eq:enstrophy} 
that leads to finite dissipation, with evidence from figure \ref{fig:nuZ}
for a numerical {\it dissipation anomaly}.

The definition of a numerical {\it dissipation anomaly} in this paper is convergence of this 
time-integral of the energy dissipation rate $\epsilon(t)=\nu Z(t)$ \citep{KerrPRF2023}. 
\vspace{-2mm}
\EQL{eq:dissanom}\Delta E_\epsilon =\int_0^{\Delta T_\epsilon} \epsilon\,dt>0\,,
\quad\mbox{with}~\Delta T_\epsilon~\mbox{a finite time }\,.\EN
Note that 
$Z(t)=V_\ell\Omega_1^2(t)={\cal O}^2_{V1}$ \eqref{eq:enstrophy},
where $\Omega_1$ is the standard first Sobolev vorticity moment \eqref{eq:Omegam}.
Unforced trefoil knot calculations \citep{KerrJFMR2018,KerrPRF2023}, both of which used 
algebraic initial vorticity profiles, have already found 
convergent $\Delta E_\epsilon$ over modest (factor of 4-8) ratios of $\nu$, 
but without corroborating analysis for
turbulent scaling such as wavenumber spectra.

In answering the finite $\Delta E_\epsilon$ \eqref{eq:dissanom} question, 
these secondary issues are also addressed.
\ITM\item What makes these calculations different than the majority of existing
vortex reconnnection calculations \citep{YaoHussainARFM2022}?
\item To achieve finite $\Delta E_\epsilon$ as $\nu$ decreases, the domain size $(2\ell\pi)^3$
must be increased. What is the dependence of $\ell$ upon $\nu$ and is this compatible with 
existing mathematics? The appendices describe the new mathematics.
\item If finite $\Delta E_\epsilon$ is achieved, is there evidence for any Kolmogorov
scaling that underlies that? This will be shown using the temporal evolution of
the enstrophy spectra.
\ITN


The calculations used to determine $\Delta E_\epsilon$ incorporate two elements from 
two recent sets of unforced trefoil knot calculations \citep{KerrJFMR2018,KerrPRF2023}, 
both of which used algebraic initial vorticity profiles and had modest 
$\Delta E_\epsilon$, but without corroborating analysis for turbulent scaling such as
as wavenumber spectra.  

The first element from \cite{KerrJFMR2018} is that in order to achieve greater 
enstrophy growth and numerical convergent dissipation rates, the domain size 
${\cal L}$ must increase as the viscosity $\nu$ decreases \eqref{eq:nuc}, thus
breaking the constraint imposed by using only $(2\pi)^3$ domains. And now justified
mathematically in the appendices.

The second element is that \cite{KerrPRF2023} shows why, due to their improved stability 
properties, algebraic vorticity profiles
rather than the traditional Gaussian vorticity profiles \citep{YaoHussainARFM2022}
should be used 

This paper combines those two elements to show how a range of 
small $\nu$, moderately high Reynolds number Navier-Stokes simulations in large domains 
can, without forcing or parameterisations, get approximate temporal convergence of the 
dissipation rates $\epsilon(t)=\nu Z(t)$ at $t_\epsilon\approx 2t_x$ and consistent 
finite-time $\Delta E_\epsilon$.  \eqref{eq:dissanom} at $t\sim 1.5t_\epsilon$. 
Along with addressing the additional secondary questions listed above. 

\begin{figure}
\vspace{-0mm}
\includegraphics[scale=0.18]{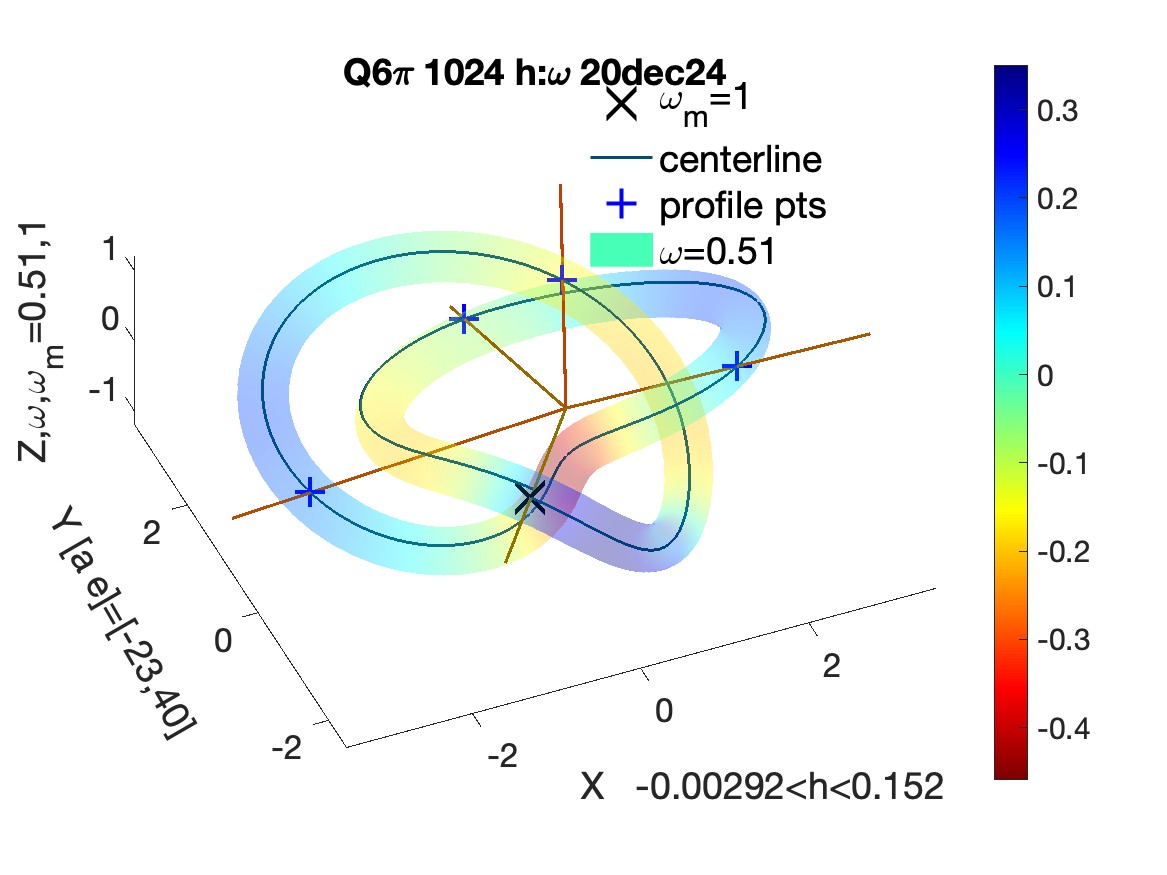}
\includegraphics[scale=0.18]{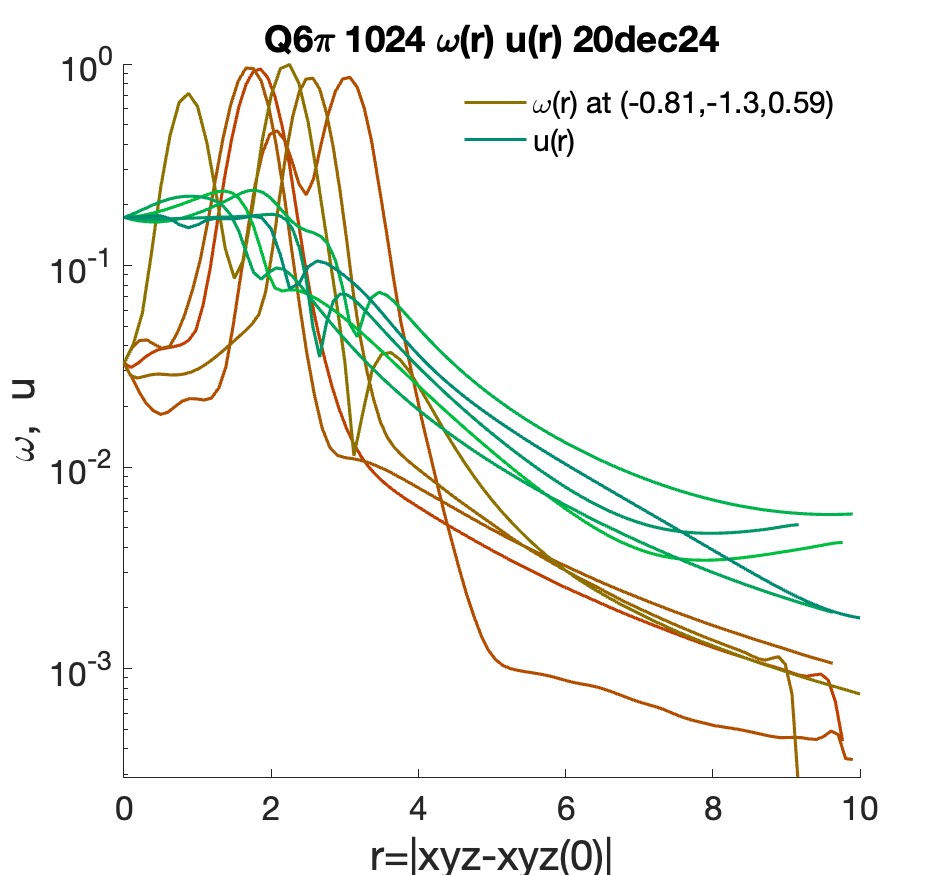}
\begin{picture}(0,0)\put(25,124){\large(a)}\put(240,60){\large(b)}\end{picture}
\vspace{-4mm}
\caption{\label{fig:T0} Perturbed trefoil initial condition. (a) $\omega=0.5$ isosurface with
$\omega_m=1$, centreline vortex seeded at $\omega_m=$ plus centreline point through which
several paths are taken. (b) Profiles of $\omega_j(r)$ taken through those points. For all the
paths $\bs_j(r)$, $\omega_j(r)\to0$ exponetially as $r$ grows. Confirming that the initial
condition is compact.
}
\end{figure}

For comparing results from several computational domains, trefoil vortex knots with
algebraic vorticity profiles are an ideal initial condition because they are compact, 
do not have inherent boundary instabilities and interesting because the usual assumptions 
of isotropy are broken by the helical initial condition. Compact means that the trefoils 
can be isolated from the boundaries as both the energy and enstrophy die off rapidly 
as $r\to\infty$ and can be redone in multiple domains.  
To demonstrate that this initial condition is compact, figure \ref{fig:T0} shows the 
$r_1=0.25$ perturbed trefoil initial condition \eqref{eq:trefoil}, with
several paths are taken from the centre of the trefoil, though its initial trajectory, 
and further out. 

Details of the stability properties of the algebraic profile on the trefoil using the 
original formulation of \cite{HowardGupta1962} have been given previously 
\citep{KerrPRF2023} Essentially what that formulation tells us is that a profile
can be unstable if the Rayleigh inflection point criteria is violated. That instability 
does not develop when algebraic profiles are used \citep{KerrPRF2023}, but can form for 
Gaussian/Lamb-Oseen profiles. However, `can form' does not necessarily mean `must form', 
and what determines that difference mathematically is whether the
initial pertubations break through ill-defined critical layers \citep{GallaySmets2020}. 
\cite{KerrPRF2023} shows that when mapping a strongly curved vortex trajectory onto 
a Cartesian mesh, the resulting field is not perfectly solenoidal and that after the 
necessary solenoidal correction is imposed, a large seed is created, one large enough to 
allow the Gaussian profile instability to grow.  This observation probably applies to 
most of the calculations discussed in a recent review \citep{YaoHussainARFM2022}. 
However, this seed does not
form if the initial vortices in a three-dimensional fluid are straight as in
\cite{ZuccoliBB2024} and \cite{Ostilloetal2021}. So there are still many situations when
use of a Gaussian profile is legitimate.

Is there an early time diagnostic, supported by mathematics, that shows whether 
increasing the domain is a viable approach? Fortunately by using the volume-integrated 
enstrophy $Z$ \eqref{eq:enstrophy} as the diagnostic, instead of the volume-averaged 
enstrophy $Z_\Omega=\Omega_1^2$ \eqref{eq:Omegam}, temporal
convergence can be achieved using $\sqrt{\nu}Z(t)$ at intermediate fixed times $t=t_x$. 
With section \ref{sec:whymath} and the appendices showing that this empirical large-domain observation 
is compatible with, through rescaling, a previously strictly $(2\pi)^3$ analytic 
result \citep{Constantin1986}.

Where have observations identified convergent energy dissipation? Historically, one means is
to use controlled experiments of turbulent flows to determine a dissipation coefficient 
$C_\epsilon$, gotten by 
comparing the measured dissipation rate $\epsilon$ to the large scale dimensional estimate 
as follows:
\EQL{eq:Cepsilon} \epsilon=C_\epsilon{\cal U}^3/{\cal L}\sim C_\epsilon r_f^2/t_{NL}^3\,,
\EN
where ${\cal U}$ and ${\cal L}$ are the large-scale velocity and length 
\citep{Vassilicos2015}.  For the trefoil one can use its diameter ${\cal L}=2r_f$ 
\eqref{eq:trefoil} and for a `turbulent' velocity ${\cal U}$, 
one can chose ${\cal U}=\|u\|_\infty(t_\epsilon)$, the maximum velocity at $t_\epsilon$, the
time with the largest dissipation rate $\epsilon$ \eqref{eq:energy} from 
figure \ref{fig:nuZ}. This is also when spectra show signs of Kolmogorov scaling in
subsection \ref{sec:trefspectra}.
Early grid turbulence and jet data showed that $C_\epsilon$ is independent of the Reynolds 
number \citep{Sreeni1984}, a conclusion that could even be applied to nonequilibrium 
turbulence as here \citep{Vassilicos2015}. However, recent results 
\citep{SchmittFPeinkeO2024} suggest that the
large-scale coefficient $C_\epsilon$ depends on the properties of the small scale 
intermittency. Those are the experimental results that are used for the comparisons 
in the inset of figure \ref{fig:nuZ}.  

The paper is organised as follows. First the equations are introduced, followed by a
discussion of why simulations in large domains 
were required, including a summary of the new mathematics qualitatively supporting the 
previous empirical conclusions. This is followed by the extended numerical results for both 
$1/(\sqrt{\nu}Z(t))^{1/2}$ and $\epsilon(t)={\nu}Z$ scaling, including evidence for a 
{\it dissipation anomaly} and additional examples of $t\lesssim t_x$, 
$\nu^{1/4}{\cal O}_{Vm}(t)$ convergence. Resolution requirements are mentioned in
section \ref{sec:IC} and discussed further in section \ref{sec:additional0p25}. 
Section \ref{sec:spectra} provides the evidence for turbulent-like spectra and finally in
section \ref{sec:3D}, some $t\lesssim t_x$ three-dimensional vortical structures are 
shown. Structures that are examples of how the $t>t_x$ range with finite dissipation 
forms. The appendix extends $(2\ell\pi)^3$, $\ell=1$ Sobolev $H^s$ analysis 
to larger $\ell>1$, $(2\ell\pi)^3$ domains, showing that as $\ell$ grows
there is relaxation of the analytic/Sobolev critical viscosities $\nu_s$ that 
enforce bounds upon the evolution of still smaller viscosities. These $\nu_s$ decrease
as $\ell$ increases, which in turn allows convergence of the 
dissipation rates $\epsilon=\nu Z$ as $\nu$ decreases. 

\subsection{Governing equations \label{sec:govern}}

This paper primarily uses the incompressible $\nu\sim Re^{-1}\neq0$ viscous 
Navier-Stokes velocity equations 
\EQL{eq:NS} \hspace{-8mm} \ppto{\bv} + ({\bv}\cdot\nabla){\bv} = -\nabla p+ 
\underbrace{\nu\Lap{\bv}}_{\rm viscous~drag}, \quad 
\nabla\cdot{\bv}=0\,,\EN
and for additional analysis, the inviscid $\nu\equiv0$ Euler equations
\EQL{eq:Euler} \hspace{-8mm} \ppto{\bu} + ({\bu}\cdot\nabla){\bu} = -\nabla p \,.\EN
This includes the equation for their difference $\bw=\bv-\bu$, which by replacing 
$\bv$ by $\bw+\bu$ throughout becomes  
\EQL{eq:weqn} \bw_t+\nu\Delta \bw =-\nu\Delta \bu 
-\underbrace{(\bw\cdot\nabla)\bw}_{B_{NL}(w,w)}
-\underbrace{(\bu\cdot\nabla)\bw}_{B_{NL}(u,w)} 
-\underbrace{(\bw\cdot\nabla)\bu}_{B_{NL}(w,u)} -\nabla p_w - \nabla p_u\,.\EN
The \cite{ConstantinFoias1988} notation $B_{NL}(u,v)=(\bu\cdot\nabla)\bv$ 
has been added to emphasize the similarities between how the three nonlinear 
terms are treated in section \ref{sec:winequ}. 

The equation for the vorticity $\bomega=\nabla\times\bv$ is
\EQL{eq:omega} \hspace{-8mm} \ppto{\bomega} + ({\bv}\cdot\nabla){\bomega} = 
({\bomega}\cdot\nabla){\bv} + \nu\triangle{\bomega},\qquad \nabla\cdot{\bomega}=0\,.\EN
and in fixed domains $V_\ell$ the vorticity moments obey this hierarchy 
\EQL{eq:Omegam} 
\Omega_m(t)=\left(V_\ell^{-1}\int_{{\cal V}_\ell}|\omega|^{2m} dV\right)^{1/2m} \EN
with $\Omega_1(t)\leq c_2\Omega_2(t)\leq\dots\leq c_m\Omega_m(t)\leq\dots\leq 
c_\infty\Omega_\infty(t)$.
This ordering represents a type of H\"older inequality where the $c_m$ are 
$V_\ell$-dependent constants. 
These moments, scaled by characteristic frequencies, have been used in 
several recent papers \citep{Kerr2013a,Kerr2013b,Donzisetal2013}.

However, the moments to be used in this paper are based on volume integration, 
${\cal O}_{Vm}(t)$ \eqref{eq:OmegamV} 
with an equivalent hierarchy ${\cal O}_{V1} \leq\dots\leq d_m {\cal O}_{Vm}(t)\leq\dots\leq
d_\infty{\cal O}_{\infty}(t)$.
Note that 
$\Omega_\infty=\|\omega\|_\infty$ is at the high end \bpurp{of both hierarchies}, and at the other end,
the volume-integrated enstrophy can be written as $Z(t)=V_\ell\Omega_1^2(t)={\cal O}_{V1}^2$.

For a compact vortex knot, the circulations $\Gamma_i$ about its vortex structures are:
\EQL{eq:Gamma} \Gamma_i=\oint_{{\cal C}_i} \bv_i\cdot \br_i 
\quad{\rm where}\quad \br_i~~
\mbox{is a closed loop about}~~{\cal C}_i\,. \EN
Dimensionally, for a vortex knot of size $r_f$ with velocity scale of $U$, the 
circulation goes as $\Gamma\sim Ur_f$. In vortex dynamics the circulation 
Reynolds number $R_\Gamma=\Gamma/\nu$ is widely used and for a vortex knot of 
size $r_f$, the nonlinear and viscous timescales are respectively:
\EQL{eq:timescales} t_{NL}=r_f^2/\Gamma\quad\text{and}\quad t_\nu=r_f^2/\nu\,.
\EN

Three local densities are used. The energy density $e=\half u^2$, the enstrophy 
density $\zeta=\omega^2$ and the helicity density $h=\bu\cdot\bomega$. 
Their budget equations, with their global integrals, are:
\EQL{eq:energy} \ppto{e}+ ({\bu}\cdot\nabla)e = -\nabla\cdot(\bu p) 
+\nu\triangle e
-\underbrace{\nu(\nabla\bu)^2}_{\epsilon={\rm dissipation}=\nu Z},~
E_e=\half\int_{{\cal V}_\ell}\bu^2dV\,;\EN
\EQL{eq:enstrophy} \ppto{\zeta}+ ({\bu}\cdot\nabla)|\bomega|^2 = 
\underbrace{2\bomega\bS\bomega}_{\zeta_p={\rm production}}
+\nu\triangle|\bomega|^2
-  \underbrace{2\nu(\nabla\bomega)^2}_{\epsilon_\omega=Z-{\rm dissipation}},~
Z=\int_{{\cal V}_\ell}\bomega^2dV\,;\EN
\EQL{eq:helicity} \ppto{h}+ ({\bu}\cdot\nabla)h = 
\underbrace{-\bomega\cdot\nabla\Pi}_{h_f=\omega-{\rm transport}}
+\underbrace{\nu\btriangle h}_{\nu-{\rm transport}} -\underbrace{
2\nu{\rm tr}(\nabla\bomega\cdot\nabla\bu^T)}_{\epsilon_h={\cal H}-{\rm dissipation}}\,,\,
{\cal H}=\int_{{\cal V}_\ell}\bu\cdot\bomega dV\,. \EN
Note that $\Pi=p-\half\bu^2\neq p_h$ and is not the pressure head 
$p_h=p+\half\bu^2$. 

In this paper the primary norms are related to the energy and enstrophy equations. Besides the
dissipation rate $\epsilon=\nu Z$ (underscore in \ref{eq:energy}), these two viscosity-based
rescalings of the enstrophy $Z$, \eqref{eq:OmegamV} will be used:
\EQL{eq:sqnuZ} \sqrt{\nu}Z(t) \quad\text{and}\quad
B_\nu(t)=1/(\sqrt{\nu}Z(t))^{1/2}=(\nu^{1/4}{\cal O}_{V1})^{-1}\,, \EN
where ${\cal O}_{V1}$ is the first volume-integrated vorticity moment 
\eqref{eq:OmegamV}.
The helicity budget is included because the helicity 
density $h$ is mapped onto the vorticity isosurfaces for potential comparisons 
with the isosurfaces of the three-fold symmetric trefoils \citep{KerrPRF2023}, 
where its budget equation is discussed in detail. 

Under $\nu\equiv0$ Euler, both $E_e$ \eqref{eq:energy} and ${\cal H}$ 
\eqref{eq:helicity}, the global helicity, are 
conserved. With the pressure gradients affecting only their local Lagrangian densities 
$e$ and $h$. In contrast, the global enstrophy $Z$ \eqref{eq:enstrophy} is not an 
invariant, but grows in figure \ref{fig:iQsnuZ} due to 
vortex stretching, as discussed in section \ref{sec:reconnect}. $E_e$ and $Z$
can be recast in terms to the first two Sobolev norms, $H^0$ and $H^1$, as defined
below \eqref{eq:uHs}. Due to vortex stretching, initially all $H^s$, $s>1$, tend to 
grow.

\subsection{Norms and inner products \label{sec:inner}}
The mathematical analysis in the appendices uses Sobolev inner products and norms. These are 
formed from the Fourier-transformed components $\hat{\bv}(\bk)$ of the velocity field $\bv(\bx)$ in a 
$(2\ell\pi)^3$ domain, with the wavenumbers $\bk=\ell^{-1}\bn$ 
written using three-dimensional integer vectors $\bn$. Giving
this definition and equation for $\hat{\bv}_{\bk}$ 
\EQL{eq:NSF} \bv(\bx)=\sum_{\bk} \hat{\bv}_{\bk} e^{i\bk\cdot\bx};\quad
\ddt\hat{\bv}_{\bk}(t)= -\nu|k|^2\hat{\bv}_{\bk}(t)-i\bP(\bk) 
\sum_{\bk'+\bk''=\bk} \hat{\bv}_{\bk'}(t)\cdot\bk'' 
\hat{\bv}_{\bk''}(t) \EN
The projection factor operator $\bP=\bI-\bk\otimes\bk/k^2$ incorporates the pressure
in Fourier space.

The inner products come from  multiplying the active equation by a test function, usually a 
velocity, in either real or Fourier space. Then integrating that over all space. This paper
uses $s$-order $\dot{H}^s=\dot{H}^s_{2\pi}$ inner products that in a 
$\ell=1$, $V_1=(2\pi)^3$ domain are determined by the power of $|k|^{2s}$ used: 
\EQL{eq:inner}\langle{u,v}\rangle_{\dot{s}}=
\int_{{\cal V}_1} |\nabla^s\bu(x)\cdot\nabla^s\bv(x)|d^3x=
(2\pi)^3\sum_{\bk}|k|^{2s}|\hat{\bu}_{\bk}(t)\cdot\overline{\hat{\bv}}_{\bk}(t)|\,.\EN
An inner product is a norm when $\bv=\bu$ with 
$\langle{u,u}\rangle_{\dot{s}}=\snm{u}{\dot{H}^s}{2}$, with
this further definition: $\snm{u}{\dot{s}}{2}:=\snm{u}{\dot{H}^s}{2}$.

The Navier-Stokes Fourier equation of the inner product of the $s=0$ norm 
$\snm{u}{0}{2}$ is particularly simple. Due to incompressibility, and by using 
integration by parts to remove the pressure, this inner product equation 
reduces to:
\EQL{eq:norm} \ddt\frac{1}{2}\sum_{\bk}|\hat{\bu}_{\bk}(t)|^2=
-\nu\sum_{\bk}|\bk|^2 |\hat{\bu}_{\bk}(t)|^2\quad\Rightarrow\quad
\ddt\frac{1}{2}\snm{u}{0}{2}=-\nu\snm{\nabla u}{0}{2}=-\nu\snm{u}{1}{2} \EN
This set of equations demonstrates the simplification provided by taking inner 
products, with the discussion in sections \ref{sec:rescalevarb} to \ref{sec:releasenu}
using the form on the right. 

The $\ell$-domain 2nd-order $s$, $H^s_\ell$ Sobolev norms of $u$, $v$ and $w=v-u$ 
\eqref{eq:weqn} can be defined using the $L^{(2)}_\ell$ and the $\dot{H}^s_\ell$ norms as 
\EQL{eq:Hs} \snm{u_\ell}{s}{2}=\snm{u}{H^s_\ell}{2}= \snm{u}{L^{(2)}_\ell}{2}+\snm{u}{\dot{H^s_\ell}}{2} \EN
where $\snm{u}{L^{(2)}_\ell}{2}=\ell^{-3}\int|u|^2 dx$ 
is the $\snm{u}{0}{2}$ inner 
product in a $(2\ell\pi)^3$ domain and the $\dot{H}^s_\ell$ norms are:
\EQL{eq:uHs} \dot{H}^s_\ell=\snm{u_\ell}{\dot{s}}{2}=\snm{u}{\dot{H^s_\ell}}{2}= (2\pi)^3
\ell^{3-2s}\sum_{k\in\dot{\bbZ}^3}|k|^{2s}|\hat{\bu}(k)|^2=
\ell^{3-2s}\snm{\nabla^s u}{}{2}\,,\EN 

In the appendices, for the higher-order $s$ norms, inequalities typically replace 
equalities like those in \eqref{eq:norm} and from this point onwards, all inner products 
are assumed to be $L_\ell^{(2)}$ and the upper $\dot{s}$ superscript will refer to the 
$\dot{H}^s$ inner products and norms.

The physical dimensions of the velocity norms, including the volume, are:
Dim$\bigl[\snm{u}{\dot{s}}{}\bigr]=\bigl[{\cal L}^{5/2-s}/T\bigr]$ 
for a chosen length scale ${\cal L}$ and time scale $T$.

\subsection{Mathematics for large $(2\ell\pi)^3$ domains \label{sec:whymath}}

A potential drawback of using periodic domains of any size is that there is a set of 
$(2\pi)^3$ mathematical bounds from applied analysis of the Navier-Stokes 
equations \citep{Constantin1986} that are largely unknown to the general
fluid dynamics community that, with further theorems, restrict the growth of the
enstrophy as $\nu$ decreases.

What was shown was that the growth all higher-order Sobolev norms under 
$\nu\neq0$ Navier-Stokes are controlled by integrals of equivalent $\nu\equiv0$ Euler norms 
as the viscosities become very small.  More precisely, from $\nu\equiv0$ 
Euler solutions $u(t)$, Analytic critical viscosities $\nu_s$ can be 
determined from time-integrals of the higher-order $s\geq5/2$, Sobolev norms $H^s(u)$.
These in turn control the growth of the equivalent $H^s(v)$ of $\nu_s\geq\nu\neq0$ 
Navier-Stokes solutions $v(t)$. 

From there, additional embedding theorems \citep{RRS2016} can then be used to show 
that the dissipation rate $\epsilon\!=\!\nu Z\!\to\!0$ \eqref{eq:energy}, 
unless there are singularities of either the Navier-Stokes or Euler equations. Simplified,
the steps are to take the bounds upon the $H^s(v)$, apply them to embedding theorems that
can bound $\|\omega\|_\infty$,
from which the vorticity moment hierarchies will bound ${\cal O}_{V1}$ \eqref{eq:OmegamV},
$\Omega_1$ \eqref{eq:Omegam} and $Z$ \eqref{eq:enstrophy}, from which
 $\epsilon\!=\!\nu Z\!\to\!0$ as $\nu\!\to\!0$ follows. 
This result is not relevant when, $\nu>\nu_s$, that is
if the viscosity is above the critical values $\nu_s$.

Those results would apply to the evolution of all initial value (Cauchy) calculations done 
in strictly $(2\pi)^3$ domains and explain why very small $\nu$, $(2\pi)^3$ 
calculations consistently fail to generate evidence for a dissipation anomaly with finite 
$\Delta E_\epsilon$ \eqref{eq:dissanom}, which is applicable to many calculations besides
those cited in \cite{YaoHussainARFM2022}. 

The appendices here modify those $(2\pi)^3$ results for larger $(2\ell\pi)^3$ domains and 
show that the suppression can be mitigated by doing small $\nu$ calculations in ever larger 
periodic domains because the \bpurp{viscosity-based $\nu_s$ bounds upon the growth of 
vorticity moments decrease as the domain parameter $\ell$ increases to accommodate decreasing $\nu$.}

The basic elements of the proof will follow a recent alternative to the classic result of 
\cite{Constantin1986} given in Chapter 9 of \cite{RRS2016}. The revisions of the
original 1986 proof fill in many gaps with essential details that did not appear 
until \cite{ConstantinFoias1988}. One of these is a robustness proof that is then
inserted where needed into the final proof. The appendices here 
then rescale the $(2\pi)^3$ results by assuming that the timescales and Reynolds numbers are
invariant as the ${\cal L}=2\ell\pi$, $({\cal L})^3$ domains are increased. Finally,
rather than just saying that certain very small critical viscosities exist, assumptions
are made that suggest their magnitudes.

\bGree{Then a better discussion in the introduction of different numerical approaches to what a dissipation anomaly is. Starting with a ARFM discussing the observation evidence for a dissipation anomaly, then other suggested explanations of how it forms. 
For example, the effect of boundary layer near singularities migrating to the interior and fluctuations from the sub-Kolmogorov scales as Eyink and Goldenfeld are promoting.}

\bGree{The dissipation anomaly question is addressed by considering the following 
dichotomy.  \cite{Vassilicos2015} quotes several sources to conclude that there is a 
consensus that whenever an energy cascade is observed, the energy dissipation rate 
$\epsilon=\nu Z$ is independent of $\nu$, 
where $Z(t)$ is the volume-integrated enstrophy (1.3). 
This contrasts with the mathematics that shows that unless there are singularities, 
$\nu\to0$ finite dissipation cannot form when the domain is fixed (Constantin 1986). 
Could the unexpected preservation of helicity during the reconnection of the 
experimental trefoil knot of Scheeler et al. (2014a) provide clues to resolving this 
dichotomy?}

\subsection{Trefoil initial condition and numerics \label{sec:IC}}

The trefoil vortex knot in this paper at $t=0$ is defined as follows:
\ITM\item[1)] $\bxi_0(\phi)=[x(\phi),y(\phi),z(\phi)$ defines the centreline
trajectory of a closed double loop over $\phi=1:4\pi$ with $a=0.5$, $w=1.5$, a 
characteristic size of $r_f=2$ and a perturbation of $r_1=0.25$.
\EQL{eq:trefoil}\begin{array}{rrl} & x(\phi)= & r(\phi)\cos(\alpha) \\
& y(\phi)= & r(\phi)\sin(\alpha) \qquad z(\phi)= a\cos(\alpha) \\
{\rm where} & r(\phi) =& r_f+r_1a\cos(\phi) +a\sin(w\phi+\phi_0)\\
{\rm and} &  \alpha=& \phi+a\cos(w\phi+\phi_0)/(wr_f) \\
{\rm with} & t_{NL}=& r_f^2/\Gamma=8 \mbox{ the nonlinear timescale } \\
{\rm and} & r_e=& (\Gamma/(\pi\omega_m))^{1/2} \mbox{ the effective radius.}
\end{array} \EN

\item[2)] The vorticity profile $|\omega(\rho)|$ uses the distance 
$\rho$ between a given mesh point $\bx$ and the
nearest point on the trajectory $\bxi_0(\phi)$: $\rho=|\bx-\bxi_0(\phi)|$. 
The profile used here is algebraic \eqref{eq:Rosenh} 
\EQL{eq:Rosenh} \omega_{\mbox{raw}}(\rho)=
\omega_o\frac{(r_o^2)^{p_r}}{(\rho^2+r_o^2)^{p_r}}\,. \EN
with a power-law of $p_r=1$, a radius of $r_o=0.25$ and a centreline vorticity before 
projection of $\omega_o\approx 1.26$.  This profile is mapped onto the Cartesian mesh,
is made incompressible as described previously \citep{KerrFDR2018},
and for algebraic profiles, the $\rho<\rho_+$ map is independent of the outer 
$\rho_+$ radius \citep{KerrPRF2023}. 
The final $\omega_o$ is chosen in each case so that the circulation is always 
$\Gamma=0.505$ \eqref{eq:Gamma} and $\|\omega(0)\|_\infty=1$, with the initial
enstrophy $Z_0=Z(t=0)\approx5.30$.
\ITN

All of the calculations in this paper use this initial condition with different domain sizes and
different mesh sizes as given in the table \ref{tab:cases}. Including the curves in figures 
\ref{fig:iQsnuZ} to \ref{fig:Om2Om4}.

The numerical method is the high-wavenumber filtered 2/3rds-dealiased pseudo-spectral code used 
previously \citep{Kerr2013a}. However, by filtering out the ubiquitous high wavenumer tail-up,
one loses a simple tool for determining whether the smallest scales are resolved 
\bpurp{by observing that tail-up.}

So alternative resolution diagnostics are needed. Comparisons between calculations with 
different resolutions is one choice, and has been applied to the $\nu=6.25\times10^{-5}$
calculations, with all of the $\Omega_m(t)$ moments for both $1024^3$ calculations, that
is for both $(4\pi)^3$ and $(6\pi)^3$, agreeing with those of the $2048^3$, $(4\pi)^3$ 
calculation. For larger Reynolds numbers, whether the resolution is adequate is determined
by these diagnostics: That the $\nu^{1/4}{\cal O}_{Vm}$ moments converge at their 
respective $t_m$ times with the lower Reynolds number calculations. This is discussed
further in section \ref{sec:additional0p25} using figure \ref{fig:issnuinfOm9}. And by their
high wavenumber spectra, as in figure \ref{fig:Q4Vk22}d, where the $Z_V(k)\sim k^{-1}$
power laws bend further into viscous exponential regimes.

On the basis of those diagnostics, the small-scales of all of large $\nu$, $1024^3$ 
calculations and the $2048^3$ calculations for the first factor of 25 decrease in $\nu$, to
$\nu=2.2\times10^{-5}$, are resolved for all times. This includes the section \ref{sec:3D}
graphics from the $\nu=3.125\times10^{-5}$ calculation with the higher vorticity isosurfaces 
during reconnnection ($t=42$ and 48) showing vortices curling around local axes.

Nonetheless, enstrophy based results for some larger Reynolds numbers with
$\nu< 2.2\times10^{-5}$ calculations are included.
For very early times the convergence of $\nu^{1/4}\Omega_\infty(t)$ and 
$\nu^{1/4}{\cal O}_{V9}(t)$ at $t_\infty=18.4$ and $t_9=19.1$ for all $\nu$ \bpurp{are shown}
in figure \ref{fig:issnuinfOm9}, providing a clear sign that these 
$\nu<2.2\times10^{-5}$ calculations are resolved for $t\lesssim20$. And
results for the enstrophy $Z(t)$ and ${\cal O}_{V1}(t)$ from three smaller $\nu$ 
calculations in large meshes are included in figures \ref{fig:iQsnuZ}b and \ref{fig:nuZ}. 
These are for $\nu=1.56\times10^{-5}$, $7.8\times10^{-6}$ and $4\times10^{-6}$. 
They are included in the sense of large-eddy simulations, with the filtering providing the 
subgrid dissipation. One justification
is that all of the previous \citep{KerrJFMR2018} results using $1024^3$ 
meshes have now been confirmed using $2048^3$ meshes. Implying that the same 
should be true for at least the new $\nu=1.56\times10^{-5}$ and $\nu=7.8\times10^{-6}$,
$2048^3$ results when re-run on $4096^3$ meshes.


\section{Vorticity moment scaling with powers of the viscosity $\nu$
\label{sec:sqnuZ}}

\begin{figure} 
\vspace{-0mm}
\includegraphics[scale=0.34]{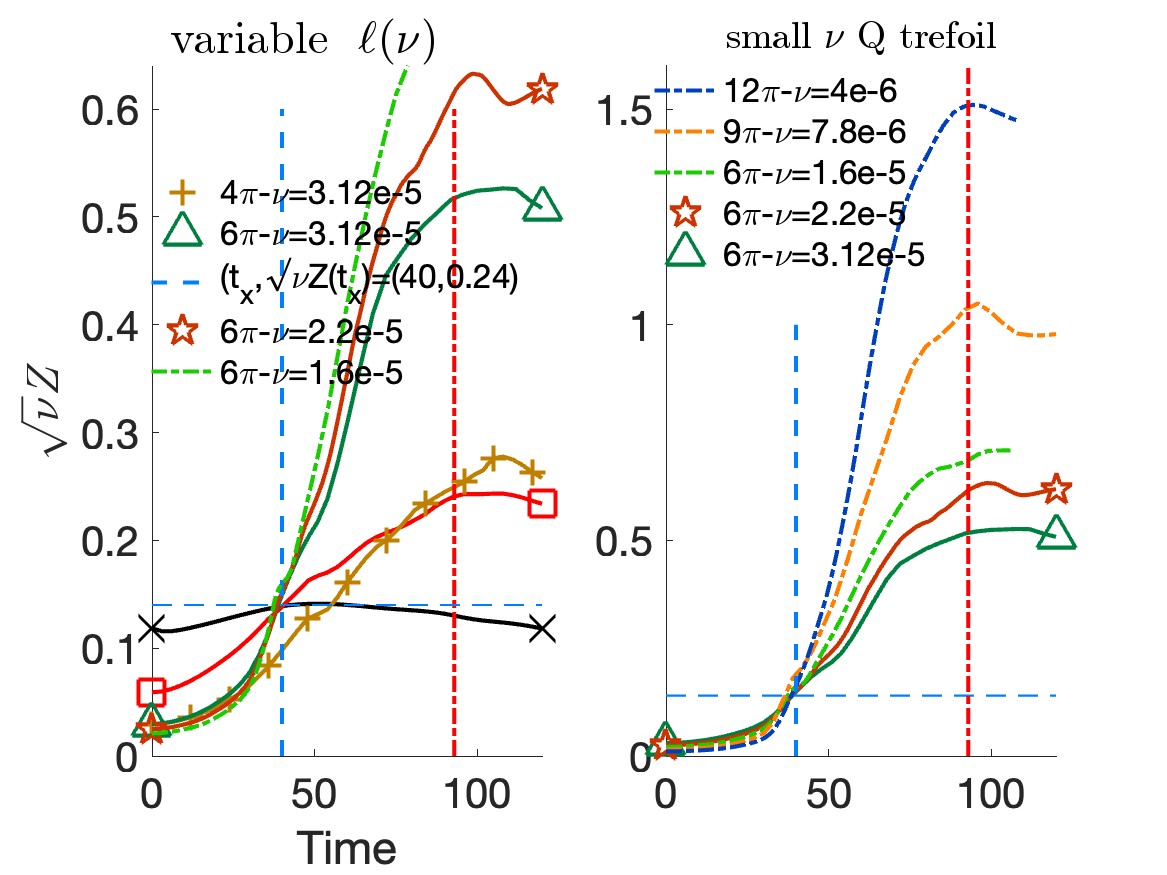} 
\begin{picture}(0,0)\put(5,274){\Large(a)}\put(370,274){\Large(b)}\end{picture}
\vspace{-4mm}
\caption{\label{fig:iQsnuZ} 
Time evolution of the {\it reconnection enstrophy} $\sqrt{\nu}Z(t)$ with all crossing at 
$t=t_x=40$ with $\sqrt{\nu}Z(t_x)=0.14$. The lower $\nu\leq3.125\times10^{-5}$ viscous cases 
were given previously \citep{KerrJFMR2018} with $t_x=40$ identified as the end of the 
reconnection phase. The line at $t_\epsilon=93$ is when plots of the dissipation 
$\epsilon=\nu Z$ approximately cross in figure \ref{fig:nuZ}. (a) The brown-$+$ curve is 
from a $\nu=3.125\times10^{-5}$, $(4\pi)^3$ calculations, the same domains as the lower
$\nu< 3.125\times10^{-5}$ viscous cases, but $\sqrt{\nu}Z(t=40)\neq0.14$. Unlike the
$\nu=3.125\times10^{-5}$ green curve that was run in a $(6\pi)^3$ domain, with 
$\sqrt{\nu}Z(40)=0.14$, plus a resolved $\nu=2.2\times10^{-5}$ case (dark-red $\star$). 
(b) Along with the two highest Reynolds
number $(6\pi)^3$, resolved calculations, $\nu=3.125\times10^{-5}$ (green-triangle) and 
$\nu=2.21\times10^{-5}$ (dark-red $\star$), three under-resolved higher Reynolds number, 
$\nu<2.21\times10^{-5}$ cases show continuing convergence of $\sqrt{\nu}Z(t)$ at 
$t=t_x=40$ as well as strong peaks at $t_\epsilon\approx93\approx 2t_x$.
}
\end{figure}

\begin{figure}
\includegraphics[scale=0.35]{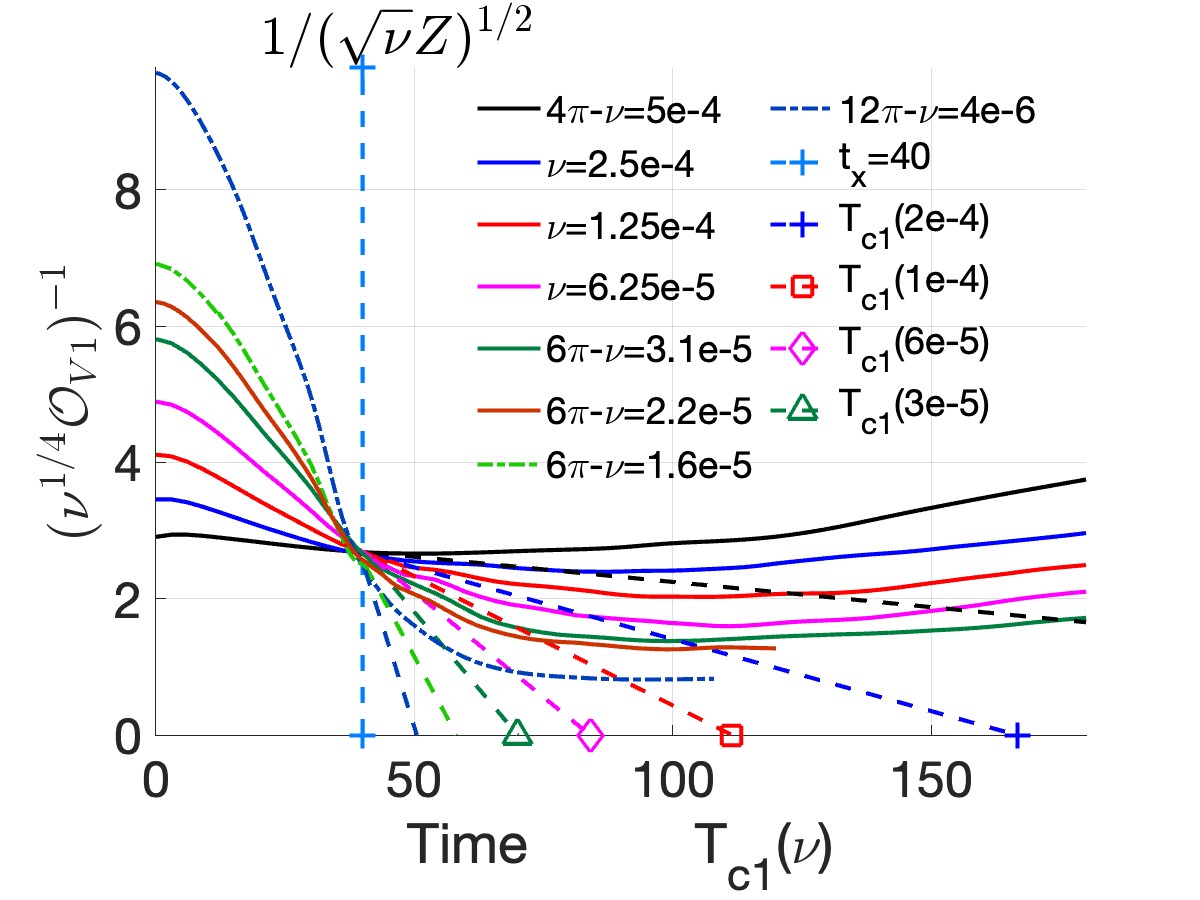} \\
\caption{\label{fig:QisnuZ} This and the next figure show further viscosity-based rescalings 
of the enstrophy $Z$. Here, using inverse $(\sqrt{\nu}Z)^{-1/2}$ scaling, rewritten as 
$(\nu^{1/4}{\cal O}_{V1})^{-1}$, shows $\nu$-independent inverse-linear convergence at 
$t=t_x=40$. To emphasise that the convergence is inverse linear, 
$t>t_x$ extensions of the $t\lesssim t_x$ behaviour are shown using long-dashed lines. 
Empirically, increases in the domain sizes $(2\ell\pi)^3$, with 
$\ell(\nu)\sim \nu^{-1/4}$, are required 
to maintain the $t\lesssim t_x$ linear scaling as $\nu$ decreases. The
$t>t_x$ linear extensions to $T_c(\nu)$ are given by \eqref{eq:TcnuDt}. 
}\end{figure}

Figure \ref{fig:iQsnuZ} introduces the unexpected (circa 2018) convergence of 
$\sqrt{\nu}Z(t)$ \eqref{eq:sqnuZ}, a {\it reconnection-enstrophy}. With $\ell>1$
larger $(2\ell\pi)^3>(2\pi)^3$, domains used to ensure temporal convergence at
a fixed time $t_x\approx40$ as the viscosity $\nu$ decreases. Using the 
${\cal O}_{V1}$ \eqref{eq:OmegamV}, \cite{KerrFDR2018} found that the best way to 
observe this convergence was to plot 
$B_\nu(t)=(\nu^{1/4}{\cal O}_{V1})^{-1}=1/(\sqrt{\nu}Z(t))^{1/2}$ 
\eqref{eq:sqnuZ}, discussed here in section \ref{sec:reconnect} where
$t_x\approx40$ is taken to be when the reconnection phase ends. As defined
by the graphics in section \ref{sec:3D}.

In 2018, it was noted that the $\nu$-independent convergence of $\sqrt{\nu}Z(t)$ 
seemed contrary to an analytic result from \cite{Constantin1986}. 
\cite{KerrJFMR2018} suggested that an analytic extension of that old result might 
exist, which is now described in the appendices. 

Two simulations with $\nu=3.1\times10^{-5}$ in figure \ref{fig:iQsnuZ} illustrate what 
happens if the domain is not increased as $\nu$ decreases.  The brown-+ curve, done in a 
$\ell=2$, $(4\pi)^3$ domain passes below $\sqrt{\nu}Z=0.15$ at $t=40$.  
The green $\triangle$ curve was done in a $\ell=3$, $(6\pi)^3$ domain and 
passes through $\sqrt{\nu}Z=0.15$ at $t=40$, which implies that 
in a $(4\pi)^3$ domain, the empirical critical viscosity for this initial condition
is $\nu_c\sim 3.1\times10^{-5}$. 

To maintain the $\sqrt{\nu}Z(t)$
convergence as the viscosity $\nu$ is decreased and the Reynolds number is increased
further, $\nu_c$ had to be decreased by increasing the domain parameter $\ell$ 
repeatedly \citep{KerrFDR2018,KerrJFMR2018}. This practice was continued several 
times, showing empirically that the critical length scale $\ell_c$ depends upon $\nu$ as, or 
critical viscosity $\nu_c$ depending upon the length scale $\ell$ as:
\EQL{eq:nuc} \ell_c\sim\nu^{-1/4}, \quad\text{or}\quad\nu_c\sim\ell^{-4}\,.\EN

Raising this question: what is origin of the dependence of the empirical 
critical viscosities $\nu_c(\ell)$ upon $\ell$ and $\nu^{-1/4}$? 
Section \ref{sec:additional0p25} discusses the ubiquitous nature of the $\nu^{1/4}$ scaling,
with the appearance of double vortex sheets in section \ref{sec:3D} seeming to underlying 
the dynamics.
However, trefoils are too complicated to be able to see the underlying dynamics clearly,
so alternative configurations such orthogonal vortices are being used, with the
preliminary conclusion summarised in section \ref{sec:summary} being:
\ITM\item The usual $\sqrt{\nu}$ scaling does appears as the double sheets are being pushed 
together. 
\item Balanced by 
$\ell\sim\nu^{-1/4}=(\sqrt{\nu})^{-1/2}$ scaling developing at their outer edges as
the sheets spread out.\ITN

\begin{figure}
\includegraphics[scale=0.35]{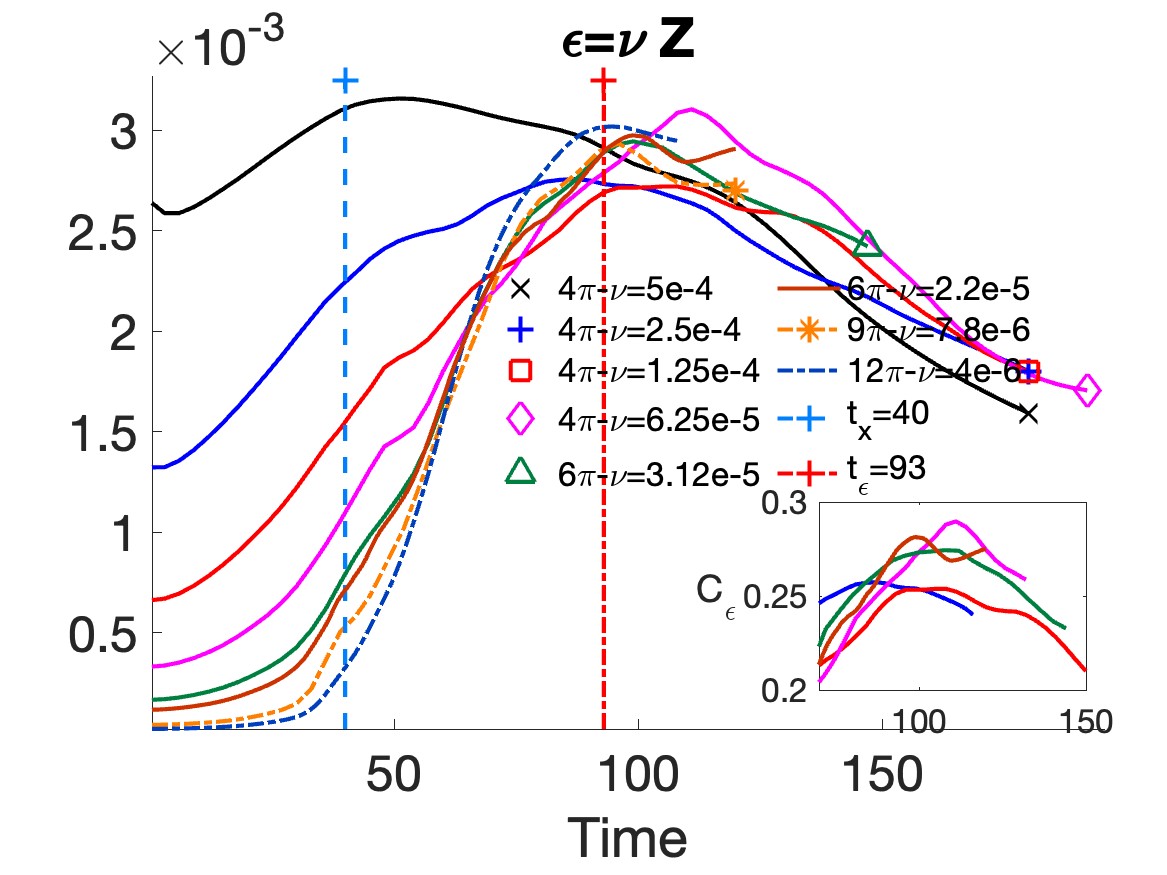}
\caption{\label{fig:nuZ} 
The dissipation rate $\epsilon=\nu Z$ for seven cases, $\nu=5\times10^{-4}$ to 
$4\times10^{-6}$, showing approximate convergence of the dissipation rates beginning at 
$t\sim70$, with peaks at $t_\epsilon\approx93\approx 2t_x$, that continues for an 
extended period. These trends are consistent with the formation of a 
{\it dissipation anomaly}, that is finite energy dissipation 
$\Delta E_\epsilon$ \eqref{eq:dissanom} in a finite time as $\nu$ decreases.
Inset: $\epsilon$ rescaled as the dissipation coefficient $C_\epsilon$ \eqref{eq:Cepsilon}
using ${\cal L}=2r_f=4$ and ${\cal U}=\|u\|_\infty(t_\epsilon=93)\approx0.34$ for
calculations whose maximum Taylor microscale Reynolds number at $t=93$ is $R_\lambda=218$.}
\end{figure}

\begin{figure} 
\includegraphics[scale=0.35]{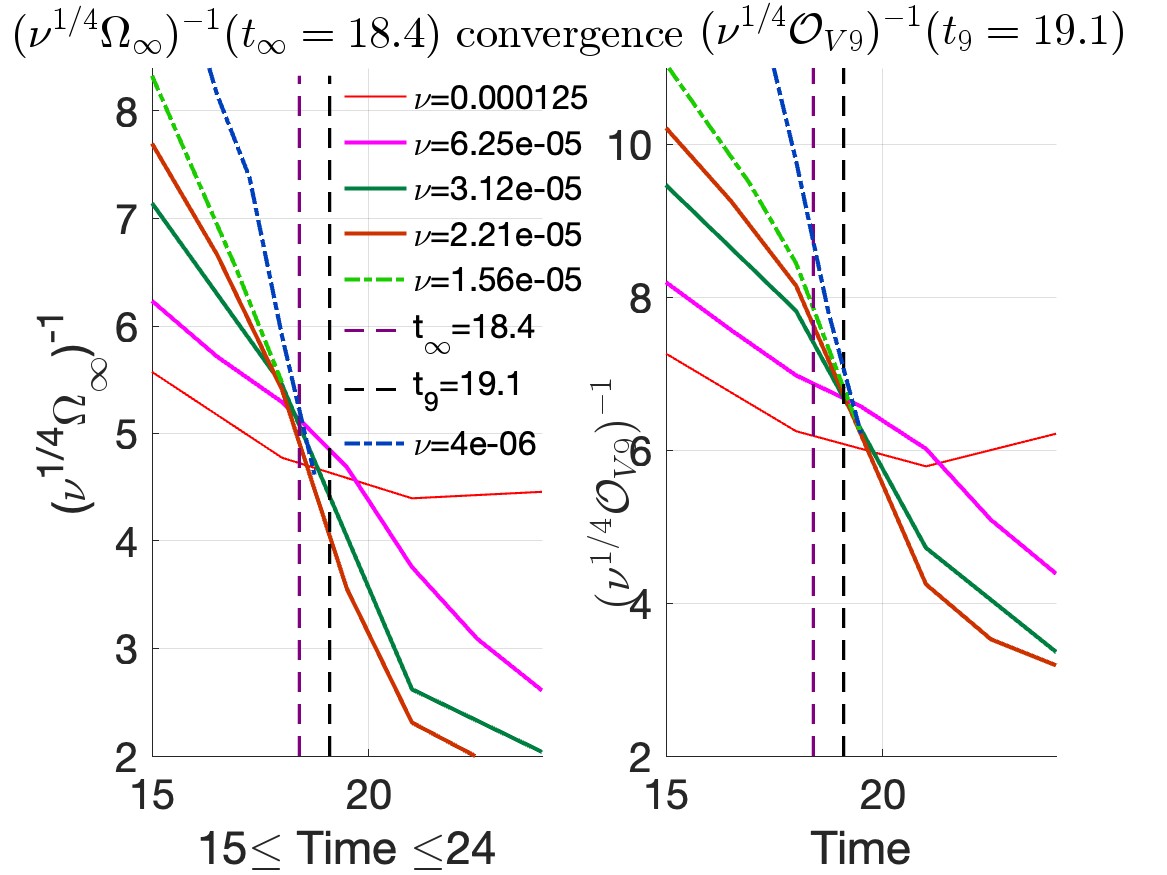}
\begin{picture}(0,0)\put(5,274){\Large(a)}\put(370,274){\Large(b)}\end{picture}
\caption{\label{fig:issnuinfOm9} (a)$(\nu^{1/4}\Omega_\infty)^{-1}$ and 
(b) $(\nu^{1/4}{\cal O}_{V9})^{-1}$. 
Recall that the end of the reconnection phase is indicated 
by when the $\sqrt{\nu}Z(t)$ converge at $t_x=40$ in figure \ref{fig:QisnuZ}. 
For $\nu\!=\!6.25\!\times\!10^{-5}$ to $\nu\!=\!4\!\times\!10^{-6}$, in (a) the 
$(\nu^{1/4}\Omega_\infty(t))^{-1}$ converge at $t_\infty\sim18.4$ and in (b) the
$(\nu^{1/4}{\cal O}_{V9})^{-1}$ converge at $t_9\!\sim\!19.1$}.
\end{figure}

\begin{figure}
\includegraphics[scale=0.35]{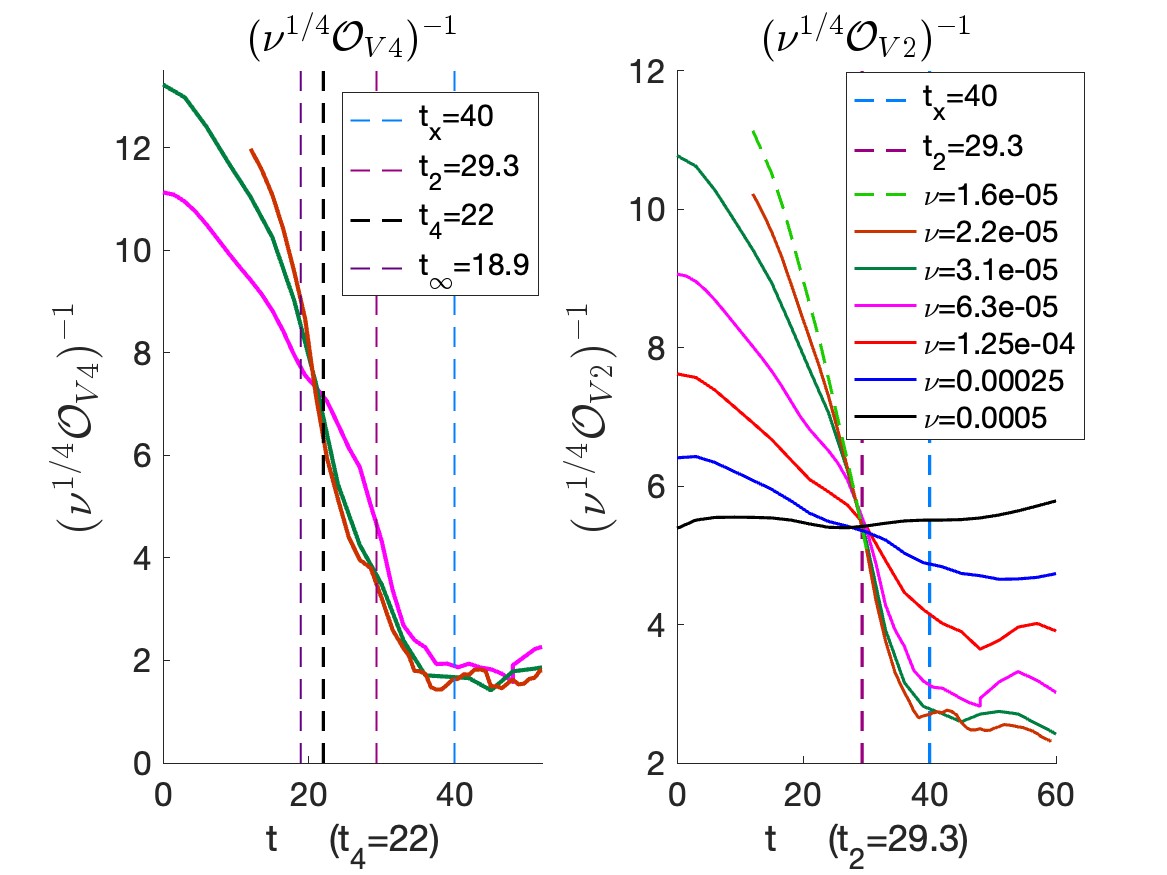}
\begin{picture}(0,0)\put(5,274){\Large(a)}\put(250,274){\Large(b)}\end{picture}
\caption{\label{fig:Om2Om4} (a) 
$\nu^{1/4}{\cal O}_{V4}$ for $t\leq50$. Recall that $t_1=t_x=40$ is the end of the
reconnection phase. Also marked are $t_2\sim29.3$, $t_4\sim22$ and $t_\infty\sim18.4$ with 
$t_x > t_2  > t_4 > t_\infty$.
(b) $\nu^{1/4}{\cal O}_{V2}$ for $t\leq60$ with $\nu=1.56\times10^{-5}$ converging at $t_2=29.3$.}
\end{figure}

Note that the convergence of $\sqrt{\nu}Z(t)$ as $\nu$ decreases is not 
convergence of the energy dissipation rate $\epsilon=\nu Z$, which comes 
later, for $t>t_x$, as the post-reconnection enstrophy growth accelerates. Unlike the 
similarity-like growth of the higher-order vorticity moments described next, similar
convergence of $\epsilon(t)=\nu Z$, around a given time should not be expected. 
Instead there is approximate convergence of $\epsilon(t)=\nu Z$ that begins at $t\sim70$ 
with peaks at $t_\epsilon\approx 2t_x$ in figure \ref{fig:nuZ}, 
with associated spectra that are discussed in section \ref{sec:spectra}.

\begin{table}
  \begin{center}
\begin{tabular}{clllllllllll}
$\tilde{\nu}(\times10^{-4})$ & 5 & 2.5 & 1.25 & 0.625 & 0.625 & 0.312 & 0.312 & 0.22 & 0.16 & 0.078 & 0.04 \\
Domain & $(4\pi)^3$ & $(4\pi)^3$ & $(4\pi)^3$ & $(4\pi)^3$ & $(6\pi)^3$ & $(6\pi)^3$ & $(6\pi)^3$ & $(6\pi)^3$ 
& $(6\pi)^3$ & $(9\pi)^3$ & $(12\pi)^3$ \\
Adequate? & good & good & good & good & good & not & good & good & good & good & good \\
Mesh & $1024^3$ & $1024^3$ & $1024^3$ & $1024^3$ & $2048^3$& $1024^3$  & $2048^3$ & $2048^3$ & 
$2048^3$ & $2048^3$ & $2048^3$ \\
Resolved? & yes & yes & yes & yes & yes & yes & yes & yes & no  & no  & no  \\
Symbols & X & $*$ & $\square$ & $\LARGE\circ$ & $\otimes$ & + & $\triangle$ & $\star$ & $--$ & $--$ & $--$ \\
Colours & black & blue & red & magenta & purple & brown & green & maroon & lime & orange & cobalt \\
\end{tabular}
\caption{\label{tab:cases} Perturbed trefoil cases parameters. Rows: $\tilde{\nu}=\nu\times10^4$.
Domain?: Is the computational domain large enough based upon 
$\ell_c$ \eqref{eq:nuc}? 
Computational Mesh, with $2048^3$ for $\nu=6.25\times10^{-5}$ to $t=48$ 
and $1024^3$ to $=192$. Resolved?: Does this mesh resolve the flow?? Symbols. Colours.}
 \end{center}
\end{table}

\subsection{Reconnection phase scaling $(\sqrt{\nu}Z)^{-1/2}=(\nu^{1/4}{\cal O}_{V1})^{-1}$ \label{sec:reconnect}}

In this subsection, figures \ref{fig:QisnuZ} and \ref{fig:nuZ} use these
two alternative viscous rescalings of $Z(t)$ \citep{KerrJFMR2018},
$B_\nu(t)=1/(\sqrt{\nu}Z(t))^{1/2}=(\nu^{1/4}{\cal O}_{V1})^{-1}$ \eqref{eq:sqnuZ} and 
$\epsilon(t)=\nu Z(t)$ \eqref{eq:energy} respectively.  Since originally introduced 
\citep{KerrFDR2018}, convergent, inverse linear $B_\nu(t)$ regimes 
have been found for the interaction of coiled vortex rings \citep{KerrJFM2018c}, 
three-fold symmetric trefoils \citep{KerrPRF2023} and additional configurations discussed 
in section \ref{sec:largescaling}.

The extensions in figure \ref{fig:QisnuZ} come from linearising $B_\nu(t)$ 
about $\nu$-independent 
$B_x=B_\nu(t_x)\approx2.67$ and extending each linear $B_\nu(t)$ over $t\lesssim t_x$ 
($30\leq t\leq40$) to define $T_c(\nu)>t_x$ as when $t>t_x$ inverse linear 
$B_\nu\bigl(T_c(\nu)\bigr)=0$. The purpose of identifying these $t>t_x$ linear extensions is to
emphasize the inverse linearity of the pre-$t_x$ $B_\nu(t)$, which suggests some type of 
self-similar process that might be found.

Specifically for $t_{30}=30$ and $t_x=40$ and using
\EQL{eq:TcnuDt} \Delta t(\nu)=B_x\frac{t_x-t_{30}}{B_\nu(t_{30})-B_x}\quad\mbox{and}\quad
T_{c1}(\nu)=t_x+\Delta t(\nu) \EN
one gets for these viscosities:
$\nu=$[0.5 0.25 0.125 0.0625 0.0312 0.0156 0.004]$\times10^{-3}$, these
$\Delta t(\nu)$= [364 127 71.5 44.1 30 18.2 10.5] and these
$T_{c1}=$[403.9 166.6 111.5 84.1 70.0 58.2 50.5].

\cite{KerrJFMR2018} used the linearity for $t\lesssim t_x$ to show how to collapse the $B_\nu(t)$ 
onto one another in a $\nu$-independent manner. This suggested that an underlying scaling process 
exists with a role for the double square-root on the viscosity. 
That is scaling using $\nu^{1/4}$ \eqref{eq:sqnuZ}. 
The next section will follow this up by extending the ${\cal O}_{V1}$ scaling to higher order 
${\cal O}_{Vm}$ and section \ref{sec:summary} will begin by mentioning other configurations
with this scaling.

\subsection{Additional $\nu^{1/4}$ scaling examples. \label{sec:additional0p25}}

Figure \ref{fig:issnuinfOm9} shows inverse $\nu^{1/4}$ scaling of $\Omega_\infty(t)$ and
${\cal O}_{V9}(t)$ \eqref{eq:OmegamV}: $\bigl(\nu^{1/4}\Omega_\infty(t)\bigr)^{-1}$ and 
$\bigl(\nu^{1/4}{\cal O}_{V9}(t)\bigr)^{-1}$, recalling that at $t=0$
$\Omega_\infty=\|\omega\|_\infty=1$ is imposed. 

Figure \ref{fig:issnuinfOm9}a shows that the $\nu$-independent temporal convergence of 
$(\nu^{1/4}\Omega_\infty(t))^{-1}$ is at $t_\infty=18.4$ and
figure \ref{fig:issnuinfOm9}b shows this for
$\bigl(\nu^{1/4}{\cal O}_{V9}(t)\bigr)^{-1}$ at $t_9=19.1$. 
For both, their temporal convergences at $t_\infty,t_9$ are best for 
$\nu=6.25\times10^{-5}$ to $4\times10^{-6}$, with the thin $\nu=1.25\times10^{-4}$ curves
at too low a Reynolds number to be \bpurp{fully} included. 
For the dot-dash $\nu<2.2\times10^{-5}$ cases, the resolution
is adequate only up to $t\sim19$, but inadequate after that for the highest-order moments.

The convergences of these large $m$, $\nu^{1/4}{\cal O}_{Vm}$ provide the earliest 
quantitative evidence that the size of the domain is affecting vorticity growth.  That is,
a sign that the local growth of $\Omega_\infty=\|\omega\|_\infty$ (and ${\cal O}_{V9}$)
is being affected by interactions across the periodic boundaries, interactions that can be
suppressed if the periodic domains increase as $\ell\sim\nu^{-1/4}$ as $\nu$ decreases. 
The $t\lesssim t_m$ evolution of all the 
$\bigl(\nu^{1/4}{\cal O}_{Vm}(t)\bigr)^{-1}$ are approximately inverse-linear, similar to how 
$(\nu^{1/4}{\cal O}_{V1}(t))^{-1}$ behaves in figure \ref{fig:QisnuZ}. 
Note that both $t_\infty$ and $t_9$ are at $\approx t_x/2$.

Can this early time evolution of $\nu^{1/4}\Omega_\infty(t)$ and 
$\nu^{1/4}{\cal O}_{V9}(t)$ tell us anything about the origins of the $\nu^{1/4}$ scaling?
For that a physical model is needed, such as understanding of the origin, extent and scaling 
of vortex sheets like those in figures \ref{fig:tr6T24oh} and \ref{fig:T30knot}. But first,
let us consider the evolution of the intermediate vorticity moments, ${\cal O}_{Vm}(t)$.

For all orders $m$ there is convergence of $\nu^{-1/4}{\cal O}_{Vm}^{-1}$ at their 
respective $t_m$, ordered as $t_\infty<\dots<t_m\dots t_1=t_x$, with $\nu$-independent 
$\sqrt{\nu}Z(t_x)$ above all others.
Figure \ref{fig:Om2Om4} shows two of these intermediate scaled, inverse moments.  
$\nu^{-1/4}{\cal O}_{V4}^{-1}$ and $\nu^{-1/4}{\cal O}_{V2}^{-1}$. For 
$\nu^{-1/4}{\cal O}_{V4}^{-1}$ from $\nu=6.25\times10^{-5}$ to $\nu=1.56\times10^{-5}$,
there is convergence at $t_4=22$ and for $\nu^{-1/4}{\cal O}_{V2}^{-1}$ 
for the same $\nu$ at $t_2=29.3$.  

One way to understand the relationship between $\nu^{1/4}\Omega_\infty$ and the 
$m<\infty$ volume-integrated moments $\nu^{1/4}{\cal O}_{Vm}(t)$, including the 
$(\sqrt{\nu}Z)^{1/2}=(\nu^{1/4}{\cal O}_{V1})^2$, is to realise that 
$\|\omega\|_\infty=\Omega_\infty$ is point-like. Because the $t_\infty=18.4$ convergence 
comes first, as $\omega$ near $\Omega_\infty(t)$ continues to grow, the region with
$\omega\sim\|\omega\|_\infty$ begins to spread out,
resulting in increases of the other ${\cal O}_{Vm}$, until one gets vortex sheets like 
those in figure \ref{fig:tr6T24oh}.


To determine whether the resolution is adequate, one common diagnostic is to follow and
compare values of $\|\omega\|_\infty=\Omega_\infty$, with figure \ref{fig:issnuinfOm9} 
showing both $\nu^{1/4}\Omega_\infty(t)$ and $\nu^{1/4}{\cal O}_{V9}(t)$ converging at 
$t_\infty=18.9$ and $t_9=19.1$ respectively for all $\nu$. Furthermore, for 
$\nu\gtrsim2.2\times10^{-5}$ and $t\geq20$, as $\nu$ decreases the inverse 
$\|\omega(t)\|_\infty$ and ${\cal O}_{V9}(t)$ continue to decrease as expected.

However for $t>20$, the smaller $\nu$, $\nu<2.2\times10^{-5}$, these highest-order
$\Omega_m$ do not decrease as would be expected, indicated that they are no longer
fully resolved and are not shown.  Leading to the conclusion that the only calculations 
that are adequately resolved for all times are those with 
$\nu\geq 2.21\times10^{-5}\approx 5\times10^{-4}/25$.



\begin{figure}
\includegraphics[scale=.33,clip=true,trim=0 0 0 0]{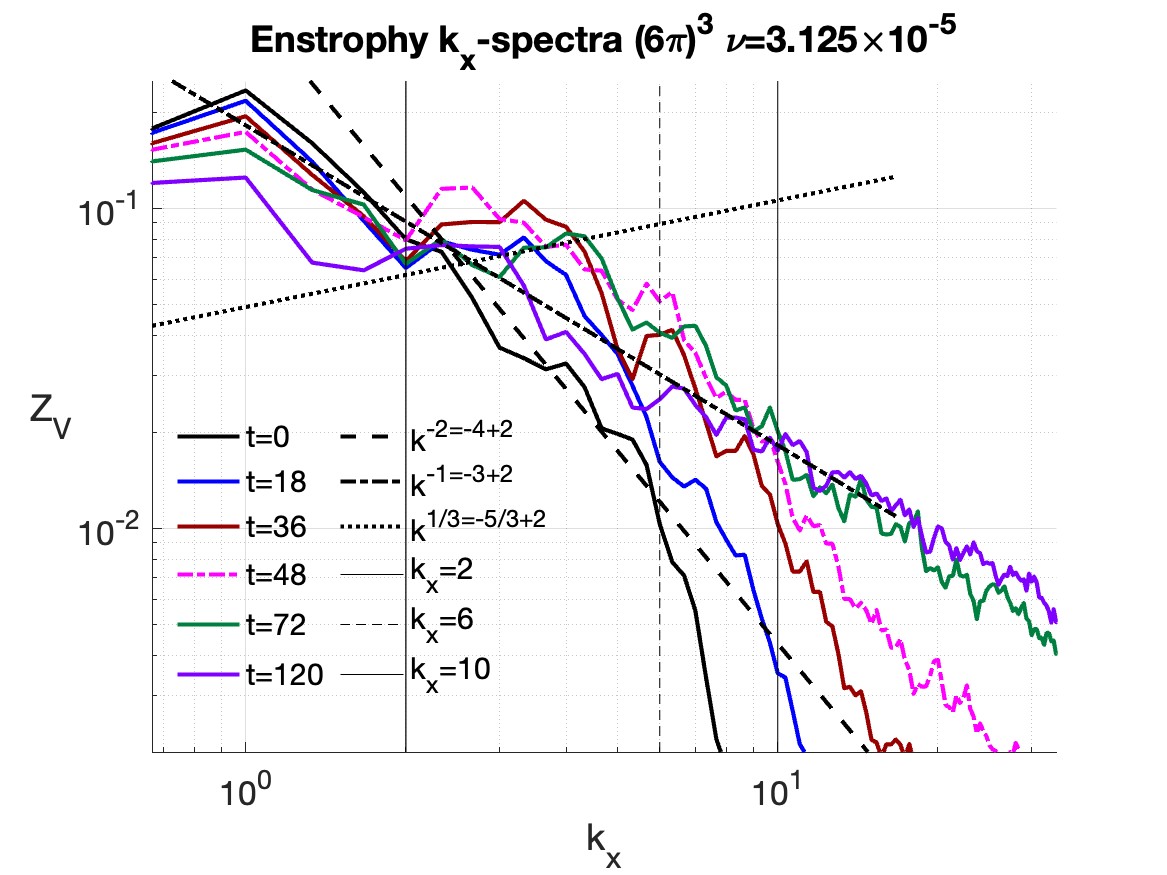}
{\Large
\begin{picture}(0,0)\put(52,208){(a)} \end{picture}}
\vspace{-4mm}
\caption{\label{fig:trQ4Vk} 
Enstrophy spectrum $Z_V(k_x)$ for seven times from the $\nu=3.125\times10^{-5}$ perturbed 
trefoil case run in a $(6\pi)^3$ domain. Times are $t=6$, 18, 36, 48, 72, 96 and 120,
plus three power laws, $k_x^{-5/3+2}=k_x^{1/3}$, $k_x^{-3+2}=k_x^{-1}$ and 
$k_x^{-4+2}=k^{-2}$. Due to the logarithmic $k_x$-scale, $k_x<2$ values are overemphasized 
compared to $k_x>2$ values. $k_x<2$ spectra generally decrease with time, excepting 
$k_x=2$ $t=48$, as small-wavenumber energy begins it transfer to higher-wavenumbers. 
$k_x\geq6$ spectra gradually
grow over time until all approximately obey $k_x^{-1}$ for all $10\geq k_x\geq30$.  }
\end{figure}

\begin{figure}
\includegraphics[scale=.30,clip=true,trim=10 0 0 0]{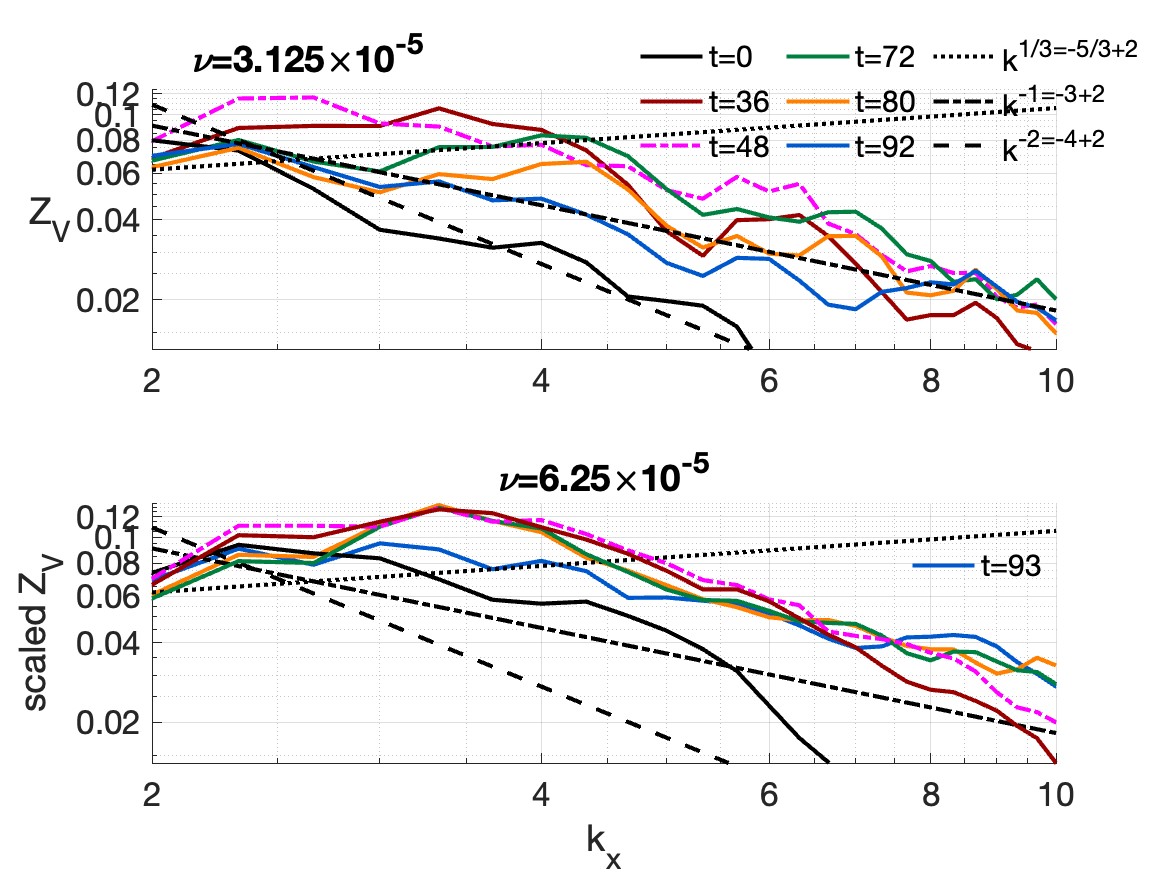}
{\Large
\begin{picture}(0,0)\put(-350,238){(a)}\put(-350,116){(b)} \end{picture}}
\vspace{-2mm}
\caption{\label{fig:Q4Vkinter} 
Comparisons of  enstrophy spectra $Z_V(k_x)$ for two intermediate resolved 
Reynolds number ($Re$) calculations of their $t=0$ and intermediate times 
($t=36$ to 92 or 93) for the wavenumber span $2<k_x<10$ to show the first step in how 
the enstrophy's form of Kolmogorov scaling, $Z_V(k_x)\sim k_x^{1/3}$, develops. 
(a) has a more thorough look at $\nu=3.125\times10^{-5}$, showing that at 
$t=48$ for $k_x\sim3$, the $Z_V(k_x)$ spectrum briefly overshoots $k_x^{1/3}$, before
becoming $Z_V(k_x)\sim k_x^{1/3}$ for $t=72$ to 80. (b)  $\nu=6.25\times10^{-5}$ shows 
that for smaller $Re$ the spectrum also overshoots $k_x^{1/3}$, but
a clear $Z_V(k_x)\sim k_x^{1/3}$ regime does not form afterwards.} 
\end{figure}
\begin{figure}
\includegraphics[scale=.30,clip=true,trim=10 0 0 0]{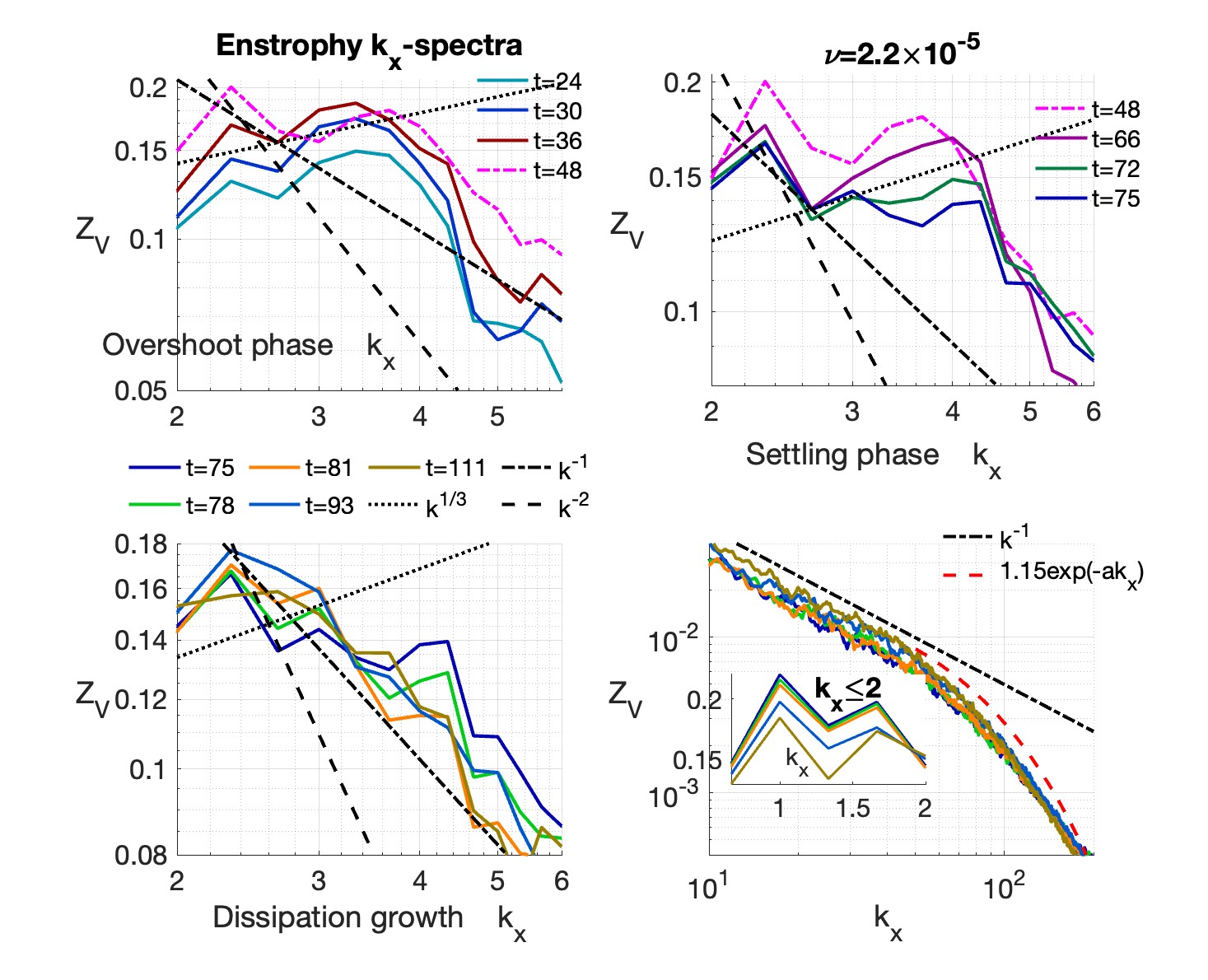}
{\Large
\begin{picture}(0,0)\put(-350,360){(a)}\put(-350,236){(b)} \put(-350,121){(c)} 
\end{picture}}
\vspace{-2mm}
\caption{\label{fig:Q4Vk22} Enstrophy spectra $Z(k_x)$ for early, intermediate and late times 
($t=24$ to 111) for the highest resolved Reynolds number ($Re$) calculation 
($\nu=2.2\times10^{-5}$) run on a $2048^3$ mesh.  (a-c) show the wavenumber span $2<k_x<6$ with
$t=48$ shown in both frames (a) and (b) and $t=75$ in (b-d).  For $t\leq36$ in (a), the spectra 
briefly overshoot $k_x^{1/3}$. (b) has the best evidence for approximately $k^{1/3}$ 
persisting from finite time span of $t=66$ to 75 for $2.67\leq k_x\leq4.33$. 
Note that the leading high-wavenumber of that $k_x^{1/3}$ span 
increases slightly with time to $t=75$, with this nascent $k_x^{1/3}$ span starting 
to decay at $t=78$ in figure \ref{fig:Q4Vk22}c. 
$t\sim70$ is roughly the time when convergence of 
the dissipation rates $\epsilon(t)$ begins in figure \ref{fig:nuZ}. Both low and very high 
wavenumbers for $t\geq75$ are shown in (d).  At high wavenumbers there is a continuation of 
the $k_x\geq5$ formation of a $k_x^{-1}$ enstrophy spectral regime from (c) 
that persists until $k_x\sim50$, with approximately 
exponential decay for $k_x>50$, indicated by the rough red-dashed fit. The inset of (d) shows
very low wavenumbers of $k_x\leq2$ with those $Z_V(k_x)$ decreasing, resuming the trend shown
in figure \ref{fig:trQ4Vk}.} \end{figure}

\section{Dissipation and spectra\label{sec:spectra}} 

The convergence properties of the $\nu^{-1/4}{\cal O}_m(t)$ scaled vorticity
moments just reported come from a phase when the dynamics is dominated by the 
emergence $h<0$ vortex sheets, which might be indicating the emergence of a 
self-similar process. Can this scaling be extended to cover convergence of the 
dissipation rates $\epsilon(t)$?

It would be surprising if the post-reconnection vorticity moment and enstrophy evolution, 
that is for $t\geq t_x\sim40$, continued this particular type of convergent scaling. Which 
$\epsilon(t)$ for  $t\geq40$ in figure \ref{fig:nuZ} shows. Nonetheless, the set of 
$\epsilon(t)=\nu Z(t)$ curves in figure \ref{fig:nuZ} do indicate that the growth of the 
post-reconnection enstrophy accelerates sufficiently to obtain approximate convergence of $\epsilon(t)$ for  $t\geq70$ up to a maxima at $t_\epsilon\sim93\sim2t_x$, which then 
persists for a period of at least $\Delta T_\epsilon\sim t_x\searrow0.5t_\epsilon$ 
at the end of these calculations. Observed both here and for the three-fold symmetric 
trefoils \citep{KerrPRF2023}. Thus providing evidence for finite 
$\Delta E_\epsilon$ \eqref{eq:dissanom} and satisfying one definition for a
{\it dissipation anomaly} with finite-time integrated energy dissipation.

Furthermore in its inset, figure \ref{fig:nuZ} shows that the dissipation constant is  
$C_\epsilon\sim0.25$ \eqref{eq:Cepsilon}, values that are consistent with the range 
of $C_\epsilon$ from a new set of
flows behind bluff bodies \cite{SchmittFPeinkeO2024}, rather than the larger $C_\epsilon$
from older wind tunnel experiments that are usually reported \citep{Vassilicos2015}.

\subsection{Trefoil spectra\label{sec:trefspectra}} 

Turbulent flows, either forced or away from boundaries, have finite energy 
dissipation rates and tend to generate energy spectra with Kolmogorov scaling. And
simulations of homogeneous turbulence, with some evidence for Kolmogorov scaling, can be
represented by forcing the flow at 
small wavenumbers, starting with \cite{Kerr1985} and now at much higher Reynolds numbers 
\citep{IshiharaetalARFM2009}. The advantage that  these statistically-steady data sets 
have is that time averages can be taken that smooth out transients.

Taking steady-state averages is not an option for the transient flows generated from
configurations of reconnecting vortices.
As reviewed by \cite{YaoHussainARFM2022}, the only Kolmogorov-like $E(k)\sim k^{-5/3}$
spectra for a reconnection simulation comes from a brief time span with maximal enstrophy 
from an anti-parallel vortex calculation.  And for the trefoil reviewed therein
\citep{YaoYangHussain2021}, its energy spectrum of $k^{-7/3}$ is much steeper than 
$k^{-5/3}$.  Furthermore, for all of those cases, the time spans over which 
those spectral scalings were observed are too short to yield a significant decrease in the 
kinetic energy. For that trefoil case, 
this has been shown to be due to the 
restrictions from using the Lamb-Oseen profile \citep{KerrPRF2023}.

Nonetheless, could a trefoil vortex knot with an algebraic initial profile, finite energy 
dissipation and temporally inhomogeneous evolution, generate energy spectra with
Kolmogorov-like scaling without taking averages?

As an alternative to determining if there is Kolmogorov scaling, without 
statistically-steady data sets, the transient enstrophy spectra $Z_V(k)$ are followed 
through the wavenumber scales to show how the enstrophy grows and is transported between 
scales. Given that the enstrophy equivalent to $E(k)\sim k^{-5/3}$ is 
$Z_V(k)\sim k^{1/3}$, further questions might include how $Z_V(k)$ might overshoot 
$k^{1/3}$, whether there are brief $Z_V(k,t)$, sub-$k^{1/3}$ periods, how long it might 
maintain $k^{1/3}$ scaling, and then how $Z_V(k)$ dissipates. 
And if a $Z_V(k)\sim k^{1/3}$ 
regime does appear for a brief period, it would only be as a bonus that could lead us 
to a new set of refined questions.

To identify the significant temporal regimes and to facilitate comparisons over several 
viscosities, only $(6\pi)^3$ perturbed trefoil calculations are shown, along with several 
plausible comparison power laws, $k^{1/3}$, $k^{-1}$ and $k^{-2}$.  
All wavenumbers are $k_x$. 
Figure \ref{fig:trQ4Vk} shows the  progression over $t=0$ to $t= 120$ of 
$Z_V(k_x,t)$ for $2/3\leq k_x \leq 33$ from the $\nu=3.125\times10^{-5}$ calculation, 
with these points
\ITM
\item $k_x<2$ spectra generally decrease with time.
\item At higher $k_x \geq 6$, there is a steady progression from the very steep $t=0$ 
spectrum to $Z_V(k_x)\sim k_x^{-1}$, $E(k_x)\sim k_x^{-3}$, spectra that are discussed 
further using figure \ref{fig:Q4Vk22}d.  
\item The $t=36$ and $t=48$ spectra overshoot the $k_x^{1/3}$ scaling. 
\item $t=48$ is marked using $-\cdot$  because this represents the transition between 
pre- and post-reconnection scaling, afterwhich $Z_V(k_x)$ spreads out until $t=72$. 
\item $k_x^{1/3}$ scaling goes through the $t=72$ curve for $2\leq k_x\leq4$ in both
figures \ref{fig:trQ4Vk} and \ref{fig:Q4Vkinter}a, with 
the full $2\leq k_x \leq 10$ behaviour in figure \ref{fig:Q4Vkinter}a.
\ITN

Figure \ref{fig:Q4Vkinter} shows the progression of $Z_V(k_x,t)$ for two viscosities
from $t=0$ to $t=92$ or 93 for $\nu=3.125\times10^{-5}$ and $\nu=6.25\times10^{-5}$ 
respectively. For both viscosities there is growth for $t<48$, $2<k_x<4$ and for $k>6$, 
as $t$
increases a $k_x^{-1}$ regime is forming. And at both $t=36$ and 48 for $k\sim 2.5-3.5$, 
overshooting of any $k_x^{1/3}$ scaling is developing. What is different between the two?
(a) For $\nu=3.125\times10^{-5}$ and $t=72$ over ($2\leq k \leq4$) there is a
$Z_V\sim k_x^{1/3}$ regime, which at $t=80$ moves to ($3\leq k \leq4.5$) 

Figure \ref{fig:Q4Vkinter}(b) for $\nu=6.25\times10^{-5}$ is shown to clarify how the 
progression of spectra appear at lower Reynolds numbers with the $k\geq2$ spectra increasing 
until $t=48$, 
including the  growing high-wavenumber $Z_V(k_x)~k_x^{-1}$ regime, which is followed by 
the $k_x\geq3$ spectra tending to decay, but not uniformly and without 
a convincing $Z_V\sim k_x^{1/3}$ regime.

These observations raise these questions:
\ITM
\item How do the spectra attain approximately $Z_V(k_x)\sim k^{1/3}$ scaling?
\item Does this spectrum persist long enough to attain finite energy dissipation 
$\Delta_\epsilon E$ \eqref{eq:dissanom}?
\ITN

To begin to answer these questions, figure \ref{fig:Q4Vk22} uses four time frames from 
the highest Reynolds number $\nu=2.2\times10^{-5}$ resolved simulation to show how a 
transient $Z_V(k_x)\sim k^{1/3}$ regime forms and how long it persists. 
Figures \ref{fig:Q4Vk22}a-c all cover $2\leq k_x\leq6$. 
\ITM
\item[$\circ$] Figure \ref{fig:Q4Vk22}a highlights the overshoot phase that ends 
between $t=36$ and 48. 
\item[$\circ$] Figure \ref{fig:Q4Vk22}b shows a settling phase with the best evidence for 
Kolmogorov-like scaling over a brief time span of $t=66$ to $t=75$ as
the enstrophy spectrum settles over $2.67\leq k_x\leq4.33$ to $Z_V(k_x)\sim k^{1/3}$ 
\item[$\circ$] This is within the times $t\sim70-120$ in figure \ref{fig:nuZ} 
when there is temporal convergence of the energy dissipation rates $\epsilon(t)$ 
\eqref{eq:energy} for several viscosities, with the maxima of the
$\epsilon(t)$ at $t_\epsilon\approx93$. 
\ITN
\ITM\item For $t\geq75$, $k_x$ is split into three parts as follows.  
\item[$\circ$] $k_x\leq2$ is in the inset of figure \ref{fig:Q4Vk22}d and is similar to the 
$t=72$ to 120 evolution for $\nu=3.125\times10^{-5}$ in figure \ref{fig:trQ4Vk}, with the 
very largest scales (smallest $k_x$) still influenced by the original, now decaying, 
trefoil structure.
\item[$\circ$]  Figure \ref{fig:Q4Vk22}c shows the influence of the high-wavenumber 
dissipation, with $Z_V(k_x)$ at the largest $k_x$, including the $k^{1/3}$ regime, 
generally decaying. 
Again within the time span of $t\sim70-120$ from figure \ref{fig:nuZ} with approximately 
convergent energy dissipation rates $\epsilon(t)$.
\item Figure \ref{fig:Q4Vk22}d shows $k_x>10$. $Z_V\sim k_x^{-1}$ from the $k_x\sim6$ span 
continues to dominate for $k_x\leq50$, followed by approximately exponential $k_x$-spectra 
for $k_x > 50$. Which is an indication that the smallest, dissipative scales of this 
calculation are adequately resolved.
\ITN

\subsection{Spectral summary \label{sec:specsum}}

Taken together, these observations show that for this unforced calculation, by increasing 
the domain, a wider range of viscosities $\nu$ and Reynolds numbers $Re$ is possible 
than those in \cite{KerrPRF2023}. A range that is wide enough to allow, beginning at 
$t\sim72$, the formation of a finite-time period with convergence of 
$\epsilon(t)=\nu Z(t)$ and spectral scaling with a precursor to Kolmogorov scaling. 
\ITM\item These are properties usually associated with turbulent flows, showing that for 
these and the earlier \cite{KerrPRF2023} calculations, these properties can form without 
the formation of a statistically-steady state. 
\item However, are these transient quasi-$k^{1/3}$ regimes
consistent with the experimental observations of Kolmogorov spectra? To conclude that
will require further analysis, including fully three-dimensional spectra plus energy flux 
and enstrophy production spectra \citep{Kerr1990}.\ITN


\section{Three-dimensional structural evolution \label{sec:3D}}

In this section two phases of the structural evolution of these flows are addressed.
One is the $t\leq t_x\approx40$ phase with $\sqrt{\nu}Z(t)$ convergence highlighted in
figure \ref{fig:QisnuZ}. The other, is the first hints of how the vortices evolve to 
generate the approximate $\nu$-independent convergence $\epsilon=\nu Z(t)$ convergence in
figure \ref{fig:nuZ}.
 
For three-fold symmetric trefoils with algebraic core profiles, 
\cite{KerrPRF2023} shows that negative helicity vortex sheets are being shed between the 
reconnecting vortices, unlike the localized braids and bridges that appear when 
Gaussian/Lamb-Oseen profiles are used, as reviewed by \citep{YaoHussainARFM2022}. 
The calculations in this paper all use the algebraic profile introduced for several 
anti-parallel Navier-Stokes and Euler calculation in 2013 
\citep{Donzisetal2013,Kerr2013a,Kerr2013b}, plus some analysis  in the \cite{KerrPRF2023} 
analysis. This profile, if isolated, would not be compact, but in a knot it is, as shown 
by figure \ref{fig:T0}. 

There are these further differences with the \cite{KerrPRF2023} trefoils. 
A different size, a different core circulation and it is perturbed. The size and 
circulation change the characteristic nonlinear timescale $t_{NL}$ \eqref{eq:timescales} 
and related times such as $t_x$, accordingly. 

\begin{figure}
\includegraphics[scale=0.225]{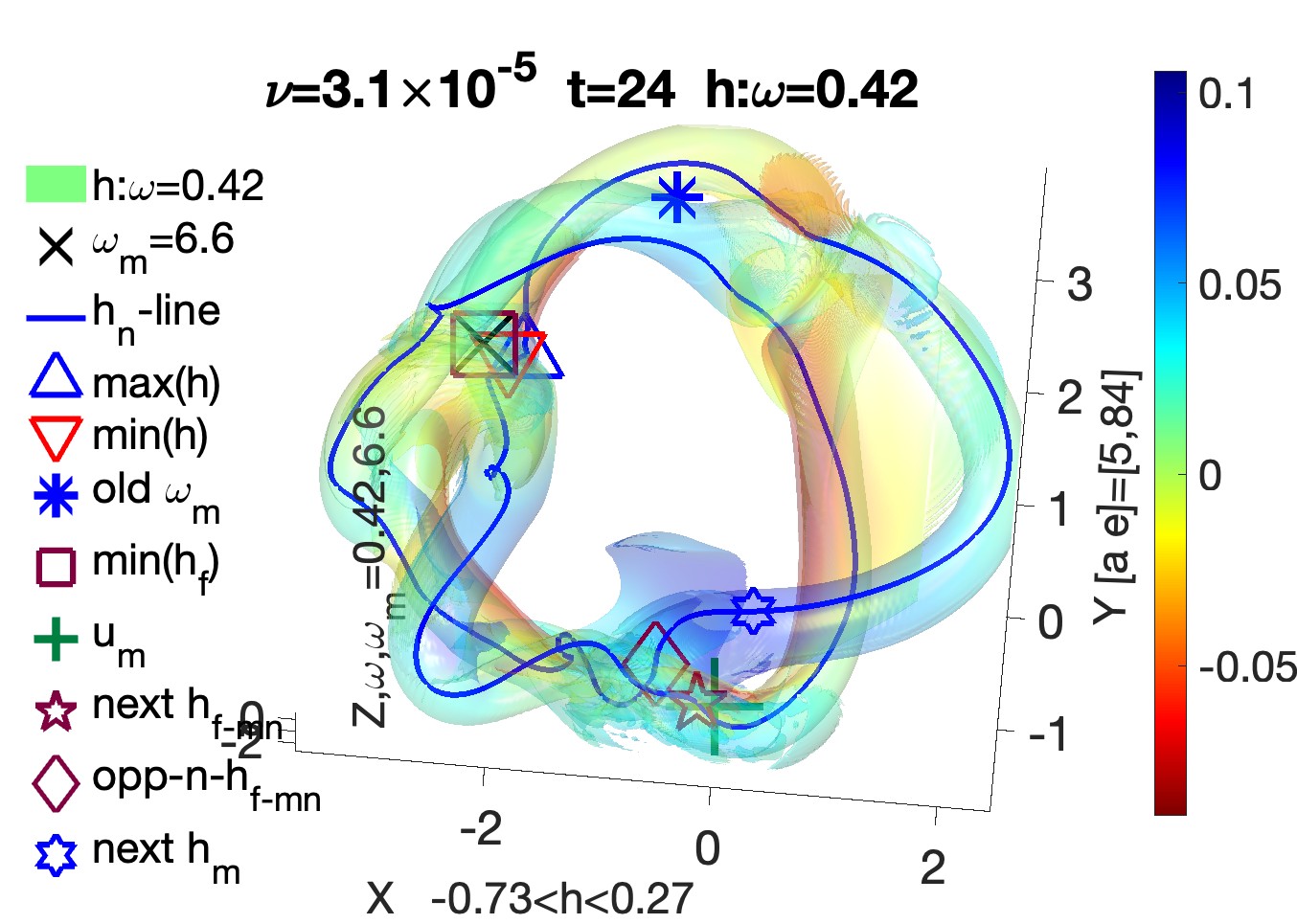} \\
\includegraphics[scale=0.225]{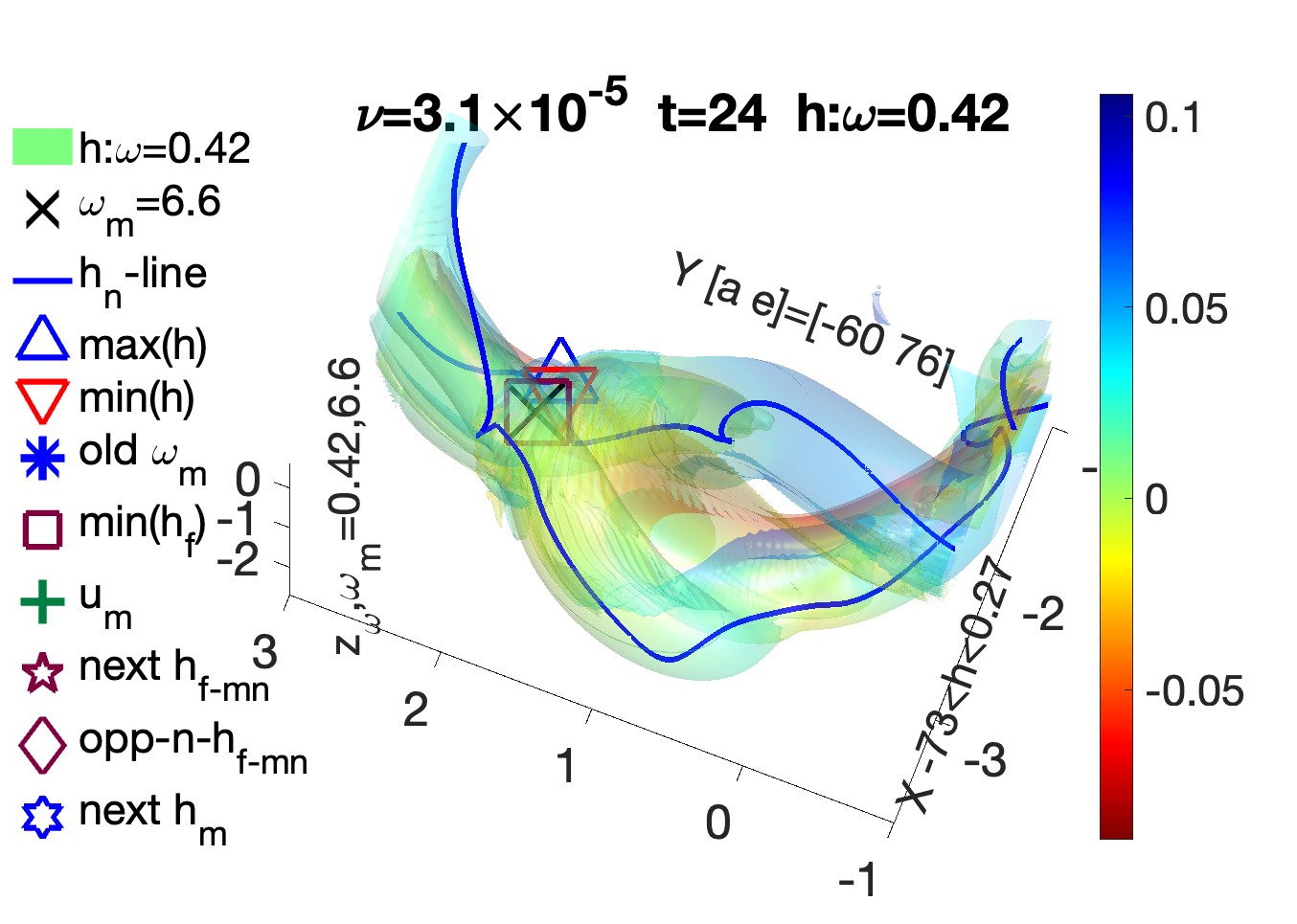}
\begin{picture}(0,0)\put(-300,496){\large(a)}\put(-300,240){\large(b)}\end{picture}
\caption{\label{fig:tr6T24oh} (a) Helicity-mapped vorticity isosurfaces from the 
$6\pi$-$\nu=3.125\times10^{-5}$ trefoil vortex knot showing the appearance of 
reddish to yellow negative helicity, vortex sheets. These are being shed off of the 
blue, predominantly positive helicity of the core of the trefoil vortex. This is 
similar to what is shown for three-fold symmetric trefoils 
\citep{KerrPRF2023} at $t=3.6 t_{NL}$.  This trefoil was perturbed so it is not 
perfectly three-fold symmetric and the evolution of the sheets at the 
three crossing locations are at different stages of sheet formation and development. 
Five primary locations are noted: $\max(\omega)=\omega_m=6.6$, $\max(h)=h_{mx}$, 
$\min(h)=h_{mn}$ and $\min(h_f)=h_{f-mn}$.  The formation of the yellowish $h\lesssim0$ 
sheet on the right began $t\gtrsim 21$ around the $t=21$ $\omega_m$ position, 
the blue $\star$ at the top. The sheet on the left is forming around the current 
$t=24$ positions of the $\omega_m$, 
$\max(h)$, $\min(h)$ and minimum of the vortical helicity flux $\min(h_f)$,
all in the upper left.  A third sheet formation might have formed at the bottom, but in 
this asymmetric case what happens instead is that the pre-existing sheets wrap around one 
another, as shown in the next two figures.
(b) Focus on left zone vortex sheets around the positions of $\omega_m$, $\max(h)$ and
$\min(h)$, rotated such the sheets are exposed. The strongest $h<0$ is coming off the outer
(bottom) trefoil leg, with a yellow $h\lesssim0$ vortex sheet between the two
legs of the trefoil.}
\end{figure}

\begin{figure}
\includegraphics[scale=0.26,clip=true,trim=50 0 0 0]{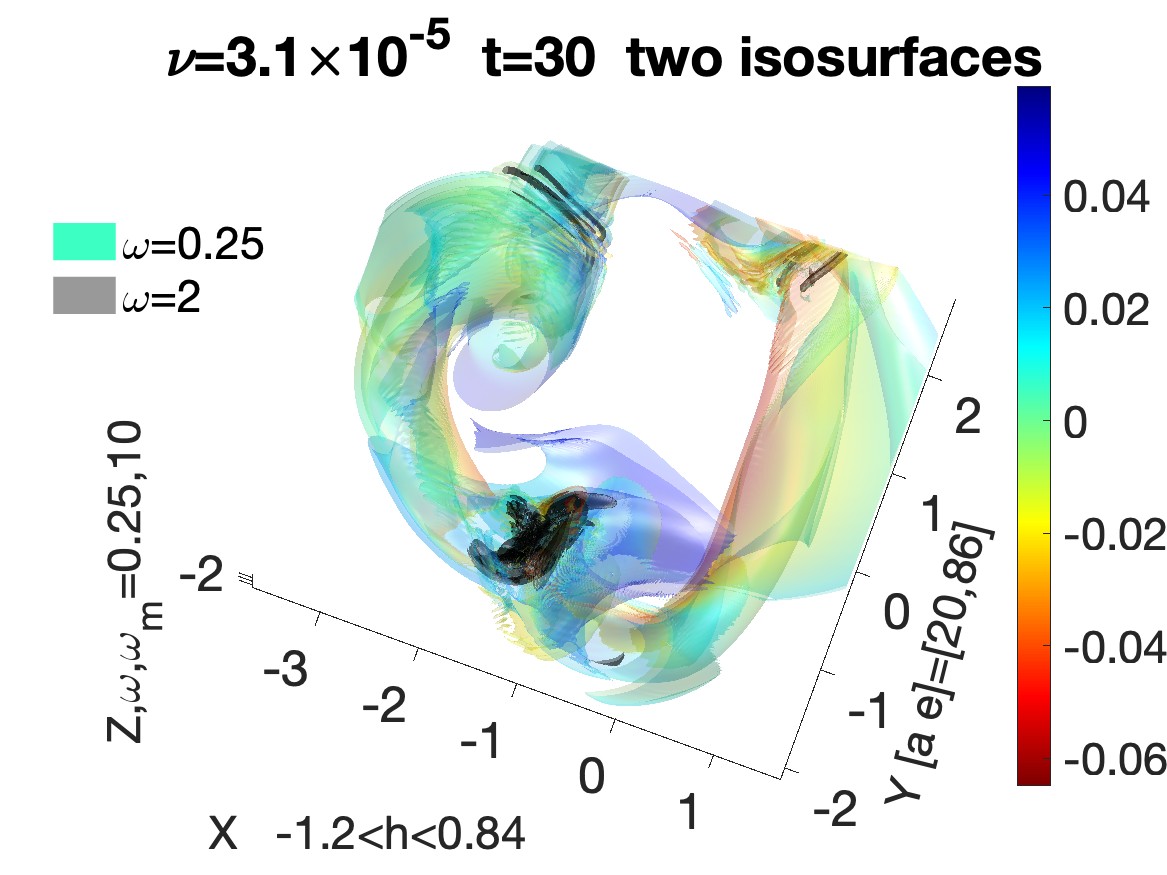} \\
\bminic{0.66}
\includegraphics[scale=0.235,clip=true,trim=50 0 0 0]{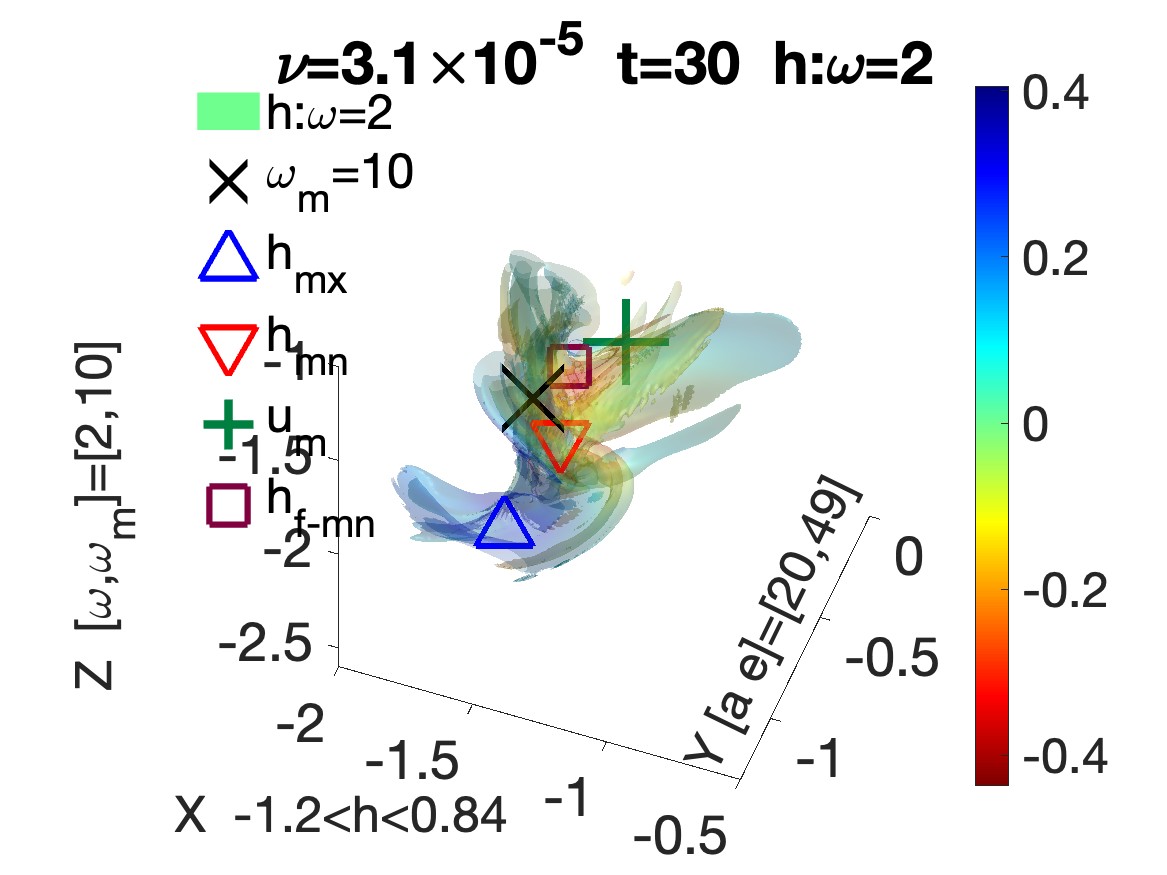}\emini
\begin{picture}(0,0)\put(-200,246){\large(a)}\put(-170,43){\large(b)}\end{picture}
\bminic{0.34} \vskip -6mm
\caption{\label{fig:T30knot} Vorticity isosurfaces at $t=30$. This and the following two 
figures focus upon the zone at the bottom from $t=24$, with the vortex sheets from figure 
\ref{fig:tr6T24oh} now winding around one another. In both frames, $\omega\!=\!2$ 
vorticity isosurfaces are shown with the maximum vorticity having increased from 
$\omega_m\!=\!6.6$ at $t\!=\!24$ to $\omega_m\!=\!10$ at $t\!=\!30$. The upper frame (a) 
shows the relationship between the primary $\omega\!=\!2$ isosurface in black and the 
helicity-mapped $\omega=0.25$ isosurfaces show the rest of the trefoil. The lower frame 
(b) uses a $\omega\!=\!2$ helicity-mapped isosurface to indicate the active
wrapping of the previously ($t=24$) formed isosurfaces, which is leading to the 
formation of an intense localised vortex knot.} \emini 
\end{figure}

\begin{figure}
\includegraphics[scale=0.29,clip=true,trim=50 0 0 0]{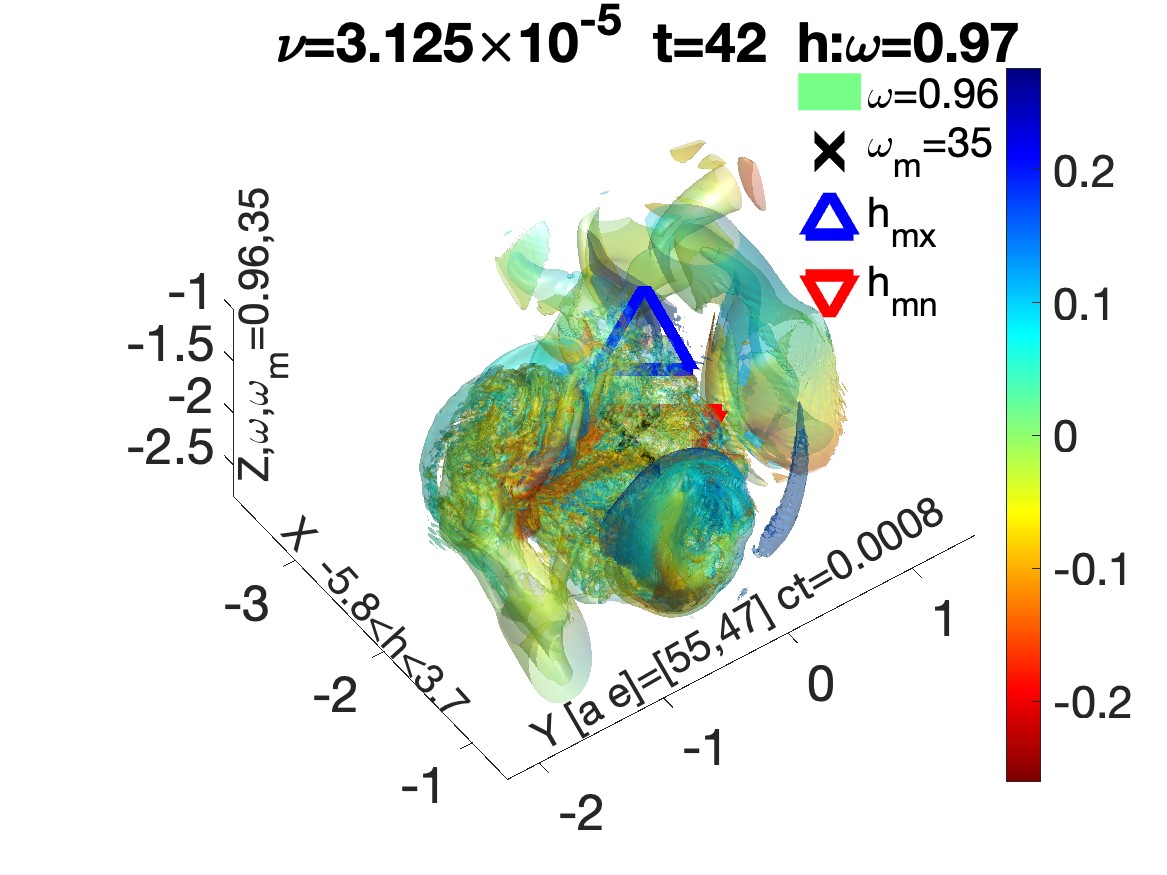} \\
\includegraphics[scale=0.29,clip=true,trim=50 0 0 0]{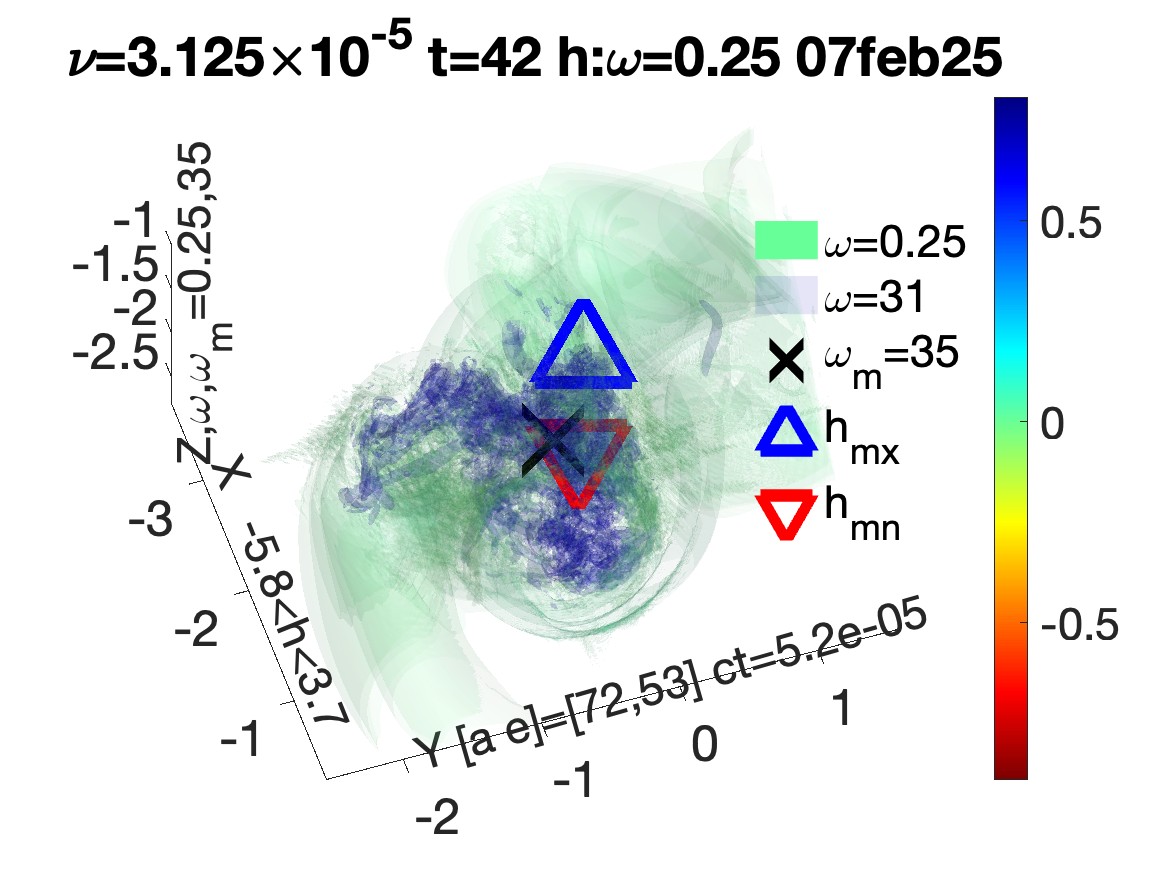}
\begin{picture}(0,0)\put(-284,500){\large(a)}\put(-290,230){\large(b)}
\end{picture}
\caption{\label{fig:T42knot} Frame (a) shows a continuation of the inner (left-side) 
$\omega=1$ vorticity isosurfaces trends begun at $t=30$ in figure \ref{fig:T30knot}b using
a $\omega=0.96$ isosurfaces. With the maximum vorticity increasing from 
$\omega_m=10$ at $t=30$ to $\omega_m=35$ at $t=42$. The more tightly wound structure on 
the right has within it the barely visible vorticity and helicity extrema, with the
maximum vorticity increasing from $\omega_m=10$ at $t=30$ to $\omega_m=35$ at $t=42$. 
This increase in $\omega_m$ is a prelude to the accelerated enstrophy production for 
$t>t_x\sim40$.  A new feature is a series of vortex tubes on the left that have 
aligned with one another. 
Frame (b) shows two additional isosurfaces
with the positions of $h_{\rm max}$, $h_{\rm min}$ and $\omega_m$ shown. 
The helicity-mapped $\omega=0.25$ isosurface 
shows how this knot is connected to the rest of trefoil and the $\omega=31$ isosurface
shows the coiled vortex structure within the wound up structure shown above.}
\end{figure}

\begin{figure}
\includegraphics[scale=0.28,clip=true,trim=50 0 0 0]{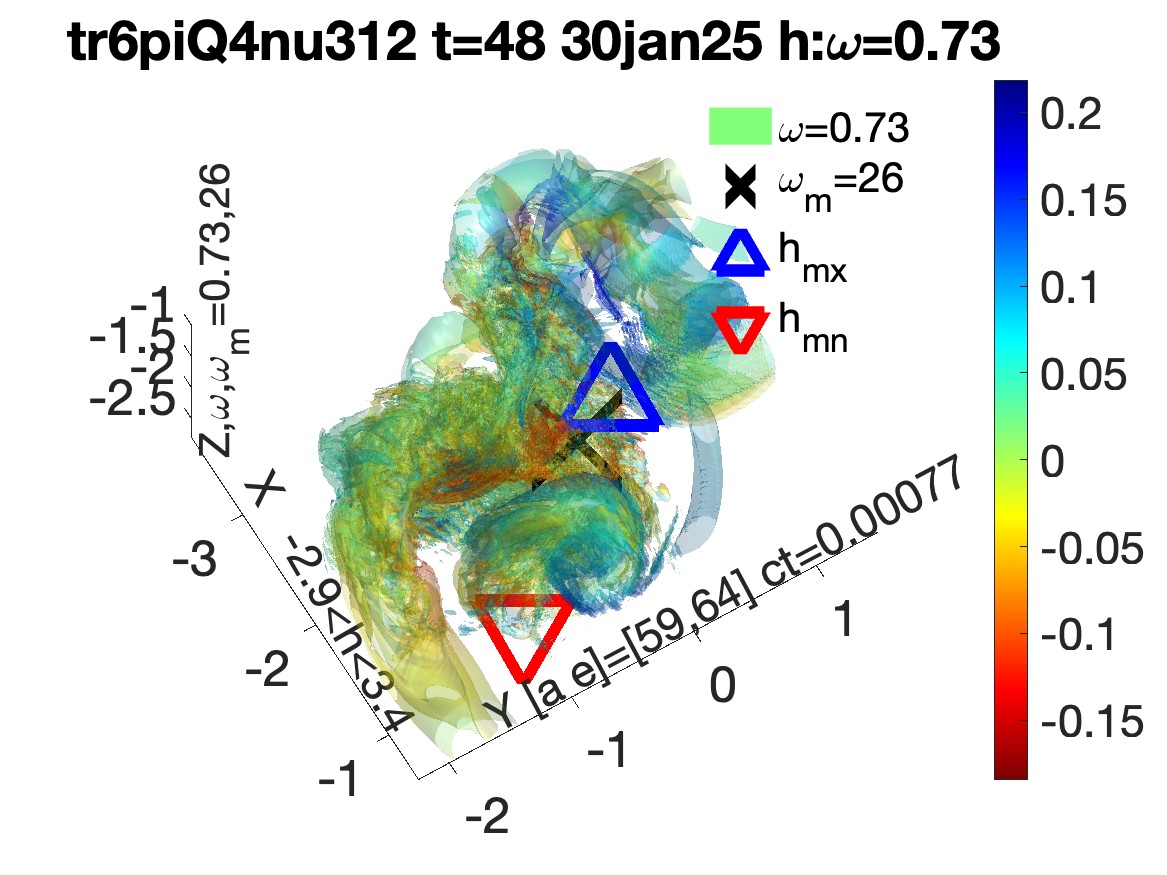}
\includegraphics[scale=0.28,clip=true,trim=50 0 110 80]{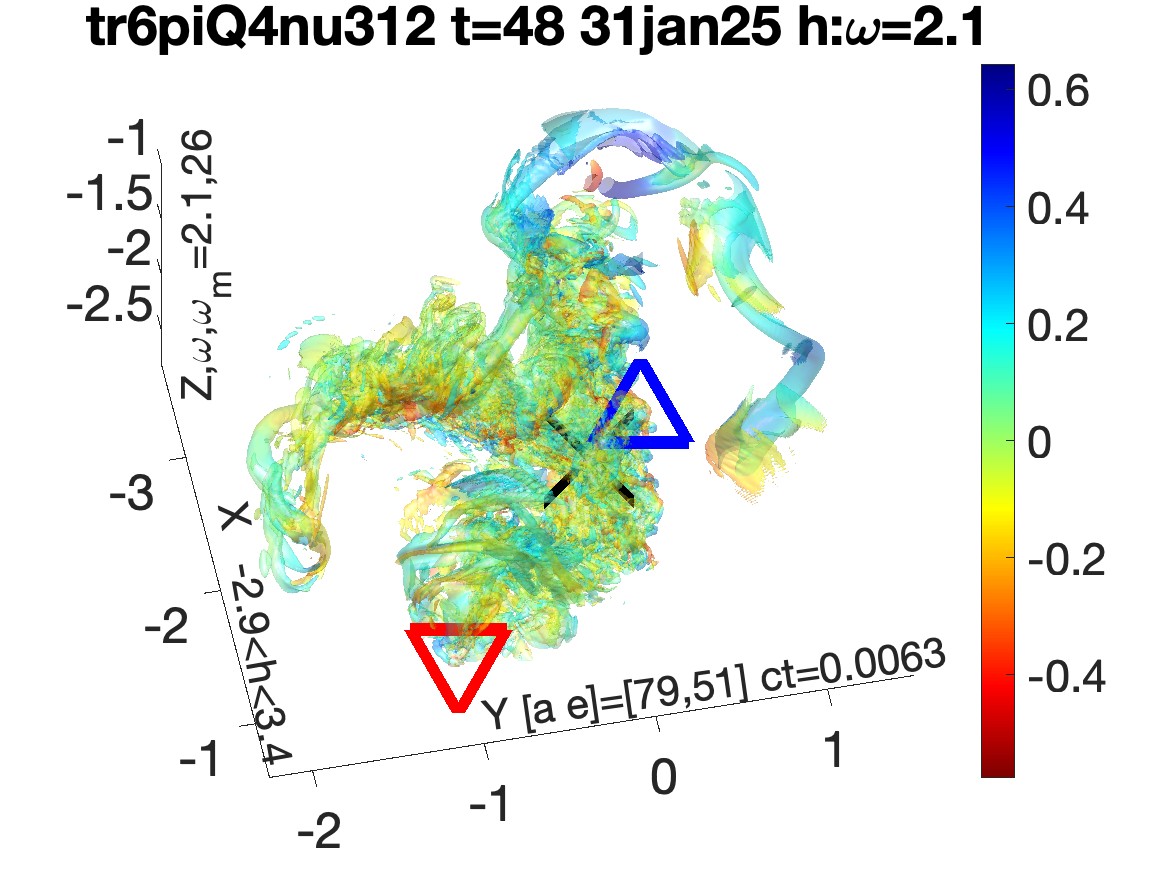}
\begin{picture}(0,0)\put(-290,354){\large(a)}\put(-355,124){\large(b)}
\end{picture}
\caption{\label{fig:T48knot} The $\omega=0.73$ isosurface in (a) is similar to that 
from $t=42$, except there is a clearer separation between the left side and wound-up 
right side. The $\omega=2.1$ isosurface in (b) makes this post-$t=42$ split more obvious 
with spirals wrapping around their respective centrelines. The dynamics of these spirals 
as it continues would be
consistent with the accelerated growth in the enstrophy that leads to the finite
dissipation $\Delta E_\epsilon$ shown in figure \ref{fig:nuZ}.}
\end{figure}

Furthermore, the perturbation changes the sequence of reconnection phase events that 
eventually lead to the production of the finite-time, finite energy dissipation in 
figure \ref{fig:nuZ}. The graphics in  figures \ref{fig:tr6T24oh}, \ref{fig:T30knot}, 
\ref{fig:T42knot} and \ref{fig:T48knot} for $t=24$, 30, 42 and 48 show the preliminary steps, but do not continue into the final dissipation phase, which will require another paper. Several isosurfaces and/or views are shown for each of these times.

The major change in how this perturbed trefoil self-reconnects comes from how the vorticity 
structures evolve at and around the three trefoil crossings, or knots. To begin, vortex 
sheets are shed from only two of the knots. These sheets then interact with one another and 
generate the vorticity maxima around the location of the third knot. And it is within 
this region that 
the $t>40=t_x$ enstrophy growth accelerates as reconnetion begins to split it apart. 

Figure \ref{fig:tr6T24oh} presents two views of the two vortex sheets that have formed by 
$t=24$.  The origin of the negative helicity sheets begins with the positions of the
minima (maximum negative values) of the $h_f$ transport term in the helicity budget
equation \eqref{eq:helicity}, which \cite{KerrPRF2023} studied in detail. 

What the local $h_f$ terms do is they push $h\!>\!0$ towards local maxima of the helicity 
and vorticity on the vortex lines, $\!h<\!0$ in the opposite direction and they flatten
the vortex tubes due to
the coincidence of the local $\min(h_f)$ with local vorticity compression. That
is local negative maxima of the enstrophy production term $\zeta_p$ in \eqref{eq:enstrophy}.

\ITM\item[$\circ$]
Now consider the maroon star at the bottom of figure \ref{fig:tr6T24oh}a with one of 
the local negative $h_f$. $h>0$ is flowing towards the local $\max(h)$, blue hexagon to
the right. 
\item[$\circ$] And $h<0$ to the left, towards the global $\min(h)<0$ at the edge of the
developing $h<0$ sheet highlighted in \ref{fig:tr6T24oh}b. 
\item[$\circ$] Where the global $\min(h_f)$ is
located there is also local $\min(\zeta_p)<0$ enstrophy compression \eqref{eq:enstrophy}, 
which is pushing that $h<0$ out into the developing sheet.
\ITN

These vortex sheets do not appear simultaneously due to the initial perturbation and the
frames of figure \ref{fig:tr6T24oh} at $t=24$ can tell us about the origins of that later 
progression. The details are:
\ITM
\item Figure \ref{fig:tr6T24oh}a is a tilted $x-y$ plan view that indicates all three 
zones of activity while also highlighting the first yellowish $h\leq0$ vortex sheet on 
the right.  This sheet originated from the blue star near the top,  the position of 
$\omega_m$ at $t=21$,  with its first growth shown by the orange blob to its right.
At $t=24$ this sheet is sitting between two legs of the original trefoil on the right.
\item The focus of Figure \ref{fig:tr6T24oh}b is upon the vortex sheet on the left 
in \ref{fig:tr6T24oh}a from a different perspective. This sheet is being actively 
spawned out of the the new positions of $\omega_m$, $\max(h)=h_{mx}$, $\min(h)=h_{mn}$ 
and $\min(h_f)$, the helicity flux minimum \eqref{eq:helicity}, whose importance in 
spawning vortex sheets has been recently discussed \citep{KerrPRF2023}.
\item The third location of strong activity at $t=24$ is indicated by the maroon 
$\star$ at the bottom. This location will not generate its own vortex sheet. Instead, 
starting at $t\sim30$, this becomes the location around which the two previously 
generated sheets wrap around one another.
\item Another role played by the vortex sheets as they extend over time is in how their 
unseen edges reach the periodic boundaries, which then push back against the growing 
sheets. This effect can provide us with
an empirical explanation for the origin of the observed critical domain 
size parameter $\ell_c\sim\nu^{-1/4}$ \eqref{eq:nuc}. Another example of the expanding 
sheets is given by figure 22 of Kerr (2023).
\ITN

Reconnection phase figures \ref{fig:T30knot} and \ref{fig:T42knot} at $t=30$ and 42 
show the further 
development around that bottom location. Both $t=30$ frames in figure \ref{fig:T30knot} have 
$\omega=2$ isosurfaces. \ref{fig:T30knot}a is dominated by a temporally extended, 
smaller vorticity, $\omega=0.25$, isosurface that is a version of the $t=24$ 
helicity-mapped vortex sheets from figure \ref{fig:tr6T24oh}a and is overlaid with 
several higher vorticity gray (single colour) isosurfaces. Two small, remnant gray surfaces 
at the top of \ref{fig:T30knot}a show where the knots that created the two $t\leq24$ 
surfaces were, with a major convoluted surface at the bottom of figure \ref{fig:T30knot}a 
being where those two $t=24$ isosurfaces are now interacting.

Figure \ref{fig:T30knot}b maps the helicity onto that higher vorticity isosurface at the 
bottom, generating a structure that shows how the already generated vortex sheets are 
being pulled together, then starting to wrap around one another. All the primary 
vorticity and helicity-related extrema are within this structure.

By $t=42$, the outer vortex sheets from $t=30$ have begun to dissipate, leaving behind in 
figure \ref{fig:T42knot}a a $\omega=0.96$ local jellyroll of vortex sheets. 
This jellyroll might be separating from the rest of the trefoil, with the $\omega=0.25$ 
isosurface in \ref{fig:T42knot}b showing some remaining connection to that outer 
structure of the trefoil. A high threshold $\omega=31$ isosurface is included in 
figure \ref{fig:T42knot}b to demonstrate that distinct coiled vortex structures exist 
within the $\omega=0.96$ isosurfaces above. 

The outer $\omega=0.73$ isosurfaces at $t=48$ in figure \ref{fig:T48knot}a show that the 
jellyroll separation noted at $t=42$ can now be describing as a splitting, with the inner 
$\omega=2.1$ isosurfaces in figure \ref{fig:T48knot}a showing spiraling vortices around 
the two halves of the split knot.  This suggests how for each of these halves enstrophy 
growth can accelerate sufficiently to generate the convergent dissipation rates in
figure \ref{fig:nuZ}, and could also be the source of the transient span of mid-wavenumber 
Kolmogorov scaling $Z_V(k)\sim k^{1/3}$ regimes in figure \ref{fig:Q4Vk22}b. One mechanism
that might explain the origin of this enstrophy increase and beginnings of Kolmogorov
scaling is the spiral vortex model \citep{Lundgren1982}.  Showing how this wrapping, 
or an equivalent small-scale 
dynamics, continues to later times will be the topic of another paper.  

\section{Summary of using large $(2\ell\pi)^3$ domains. \label{sec:whylarge}}

This paper has focused upon two aspects of the large domain, higher Reynolds number 
trefoil calculations that were not fully covered by the earlier trefoil calculations of 
\cite{KerrJFMR2018} and \cite{KerrPRF2023}. 
The first aspect, in sections \ref{sec:reconnect} to \ref{sec:3D}, extends the 
numerical analysis of the perturbed trefoil calculations from before \citep{KerrJFMR2018},
with later times and higher resolution for most, plus one new calculation. 

These extensions begin with enlargements of the computational domains 
at mid- to late-times so as to maintain the observed temporal convergence of 
$\sqrt{\nu}Z(t)$ at $t=t_x\approx40$ in figure \ref{fig:iQsnuZ}, and then
approximate convergence of the dissipation rate $\epsilon=\nu Z$ for 
$t\sim t_\epsilon=93\sim2t_x$ in figure \ref{fig:nuZ}.  This enlargement included 
the smallest viscosities/highest Reynolds numbers calculations, all of which are indicated 
by dot-dash lines and are somewhat under-resolved, These are
$\nu=1.56\times10^{-5}$, $7.8\times10^{-6}$, and $4\times10^{-6}$. 
Previously $t_x\approx40$ was identified from vortex structures as the end of the 
reconnection phase \citep{KerrFDR2018}, and then as the transition from  the inverse-linear 
$\bigl(\sqrt{\nu}Z(t)\bigr)^{-1/2}$ behaviour \citep{KerrJFMR2018}, as in figure 
\ref{fig:QisnuZ}, to accelerated enstrophy growth that leads to convergence of the 
energy dissipation rates $\epsilon=\nu Z(t)$ in figure \ref{fig:nuZ}. That $t_x\approx40$ 
is the time when physical space reconnection begins is shown here using graphics at 
$t=42$ and 48 in figures \ref{fig:T42knot} and \ref{fig:T48knot}.

Figure \ref{fig:nuZ} shows that the convergence of $\epsilon(t)$ begins at $t\sim70$ and 
persists for approximately $\Delta T\sim t_x$, which is long enough that when 
temporally integrated yields approximately $\nu$-invariant finite 
$\Delta E_\epsilon$ \eqref{eq:dissanom}, satisfying
one definition for a {\it dissipation anomaly}. This $\epsilon(t)=\nu Z=\nu{\cal O}_{V1}^2$ 
convergence can be obtained over the full factor of 125 decrease in $\nu$ and is consistent 
with these being turbulent flows.  
However, convergence of the higher-order $m>1$ ${\cal O}_{Vm}(t)$ 
moments at these later times can only be achieved over a factor of 25 in the
Reynolds number. Figure \ref{fig:Q4Vk22} shows that for these mid-range Reynolds numbers,
the mid-wavenumber spectral slopes are Kolmogorov-like, providing further evidence for the 
appearance of classical turbulence.

\subsection{Supporting numerical analysis and mathematical scaling for $(2\ell\pi)^3$ domains.  \label{sec:largescaling}}

The second aspect is providing new quantitative numerical analysis and mathematical scaling
to explain why the computational domains need to be increased.  Because they are compact, 
trefoil vortex knots have been ideal for showing that the $B_{\nu}(t)$ \eqref{eq:sqnuZ} scaling 
can be retained as $\nu$ decreases by increasing the domain size, leading to
this question: What controls the domain dependence of the enstrophy
suppressing critical viscosities? 

Previously, to maintain the temporal convergence at $t_x$, the domain size 
${\cal L}=2\ell\pi$ had to be empirically increased as $\nu$ decreased 
\citep{KerrFDR2018,KerrJFMR2018}.  Although an estimate of that dependence was not 
given. Here, with a wider range of $\nu$, especially for the convergence of 
$(\nu^{1/4}\Omega_\infty)^{-1}$ at $t_\infty\approx18.4$, plus the consistency of 
the convergence of each ($t<t_x$) $(\nu^{1/4}{\cal O}_{Vm}(t))^{-1}$, 
it is clear that the increase is $\ell\sim\nu^{-1/4}$. With the structural evidence 
pointing to the growth of vortex sheets as in figures \ref{fig:tr6T24oh} and
\ref{fig:T30knot}a as the source of the limiting factor. 
Overall, combining the requirements at large scales with $\ell\sim\nu^{-1/4}$, with
those at small scales going as $\eta\sim\nu^{3/4}$, the range of scales 
should increase as $\ell/\eta\sim\nu^{-1/4}/\nu^{3/4}\sim \nu^{-1}$.

A more rigorous supporting result is given in the appendices. This provides an extremal 
mathematical justification for why increasing $\ell$ is needed.  The method consists of 
determining domain size dependent mathematical critical viscosities $\nu_s(\ell)$ as $\ell$ 
increases. These analytic/Sobolev $\nu_s(\ell)$ are found by
rescaling the high-order $s>5/2$ Sobolev analysis for $\ell=1$, $V=(2\pi)^3$ into larger 
$V_\ell=(2\ell\pi)^3$ domains. With some $(2\pi)^3$ estimates of the $\nu_s$ made from the
Euler norms in appendix \ref{sec:usewinequ}, before they are transformed into the original
domain in appendix \ref{sec:releasenu}. 

Is there mathematics that could bring the two determinations of the critical viscosities 
more in line with one another? Currently we know that the $\nu_s(\ell)$ should be, 
and are, orders of magnitude smaller than the empirical $\nu_c$. However,
it is believed by many that $\ell\to\infty$ limiting behavior can be determined by a more 
direct extension of the $(2\pi)^3$ mathematics \citep{Constantin1986}
to whole space, infinite $\bbR^3$.  Having such an extension written down explicitly might 
be helpful in determining $\nu_s(\ell)$ and bridging that gap. 

The conclusion of both approaches is that getting a {\it dissipation anomaly}, that is 
finite-time finite $\Delta E_\epsilon$, without finite-time singularities is possible only 
when the physical problem is done in whole space, that is infinite $\bbR^3$. 

\bGree{Although there is an alternative point-of-view that outer, domain-independent bounding 
length scale can be found for every compact initial condition, from which 
higher-order $\dot{H}^s_\ell$ \eqref{eq:uHs} can identify critical $\nu_s$ such that for
$\nu\leq\nu_s$, the Navier-Stokes solutions would be  bounded by $\nu\equiv0$ Euler solutions.
If this analysis exists, it needs to be published.}

\subsection{Does $(2\ell\pi)^3$ rescaling lead to turbulence? \label{sec:isturbulent}} 

Does evidence for a {\it dissipation anomaly} and Kolmogorov spectra
imply that turbulent scaling is obeyed? For example, by
following the seminal ideas of \cite{Onsager1949}, one explanation for the 
origins of finite dissipation, structure function scaling and how $k^{-5/3}$ 
Kolmogorov spectra form is to assume the existence of singularities with the velocity increments 
in space having a 1/3 H\"older exponent \citep{Eyink2003}.

However, the latest forced, very high Reynolds numbers calculations are not providing 
convincing numerical evidence because the dissipation rate $\epsilon$ 
\eqref{eq:energy} increases slowly when the Reynolds numbers
are increased \citep{IyerSreeniYeung_NLampJFM2022}. Given these difficulties, the primary
evidence in favor of the existence of finite $\Delta E_\epsilon$ in turbulent flows is 
still largely experimental \citep{Vassilicos2015,SchmittFPeinkeO2024}.

Due to these limitations upon the Navier-Stokes direct numerical simulations, alternative 
explanations for how finite-time dissipation can form have been proposed. 
These include whether there are recurrent quasi-singular events, or whether
additional sub-Kolmogorov fluctuations have roles \citep{EyinkJafari2022}. Or maybe boundary 
layer singularities \citep{LuoHou2014} might trigger significant interior three-dimensional 
dissipation rates. None of these possibilities has been demonstrated using 
three-dimensional simulations that have been run over a range of viscosities.

The references just given take the point of view that scaling of structure functions are 
critical.  Eventually an attempt to find structure functions for a compact numerical
data set should be done.  Most likely 
following a method similar to what has been applied to steady-state, homogeneous flows 
\citep{Kerretal2001}. However, like the spectral scaling in section \ref{sec:spectra}, 
those results would be time-dependent and less conclusive than desired.

The calculations here present an alternative approach to generating numerical turbulence 
where the large-scale forcing is replaced by large-scales that grow without bounds. 
Demonstrating that the incompressible Navier-Stokes, without the types of corrections 
suggested above, are capable of
generating finite-time energy dissipation and signs of Kolmogorov scaling, which is one 
manifestation of Onsager's 1/3 scaling.

\section{Summary \label{sec:summary}}

To close this paper, a number of questions are now presented, some of which of are currently
being addressed.

Is convergent scaling by $\sqrt{\nu}Z(t)=\bigl(\nu^{1/4}{\cal O}_{V1}(t)\bigr)^2$, unique to 
trefoils? In published calculations the equivalent $B_\nu(t)=(\nu^{1/4}{\cal O}_{V1})^{-1}$ 
scaling \eqref{eq:sqnuZ} has been identified 
for trefoils with different core thicknesses \citep{KerrFDR2018}, 
symmetric trefoils in a $(2\pi)^3$ periodic box 
\citep{KerrPRF2023}, and nested coiled rings \citep{KerrJFM2018c}. And in ongoing work, 
${\cal O}_{Vm}(t)$ scaling \eqref{eq:OmegamV} is being applied to the Taylor-Green vortex 
\citep{Brachetetal83} and new versions of
recent orthogonal vortices calculations \citep{Ostilloetal2021}.


What is the origin of the $\nu^{1/4}$ scaling? 
Lengths with a $\sqrt{\nu t}$ dependence are common in fluid theory. One example is how the
compression perpendicular the stretching along Burgers vortex 
is balanced by the perpendicular viscous terms. Another example is how 
some of the most refined mathematics uses \cite{Leray34} scaling with 
$\delta_\nu(t)\sim\sqrt{2\nu(T-t)}$ to restrict singularities 
of the Navier-Stokes equations \citep{Necasetal96,EscauSS03}.
However Leray scaling is about collapse to a point, not to type of sheets observed here.

Could these vortex sheets be the source of the $\nu^{1/4}$ scaling? An answer comes from
how the sheets form out of the helical knots within the trefoil configurations, here
and in \citep{KerrPRF2023}, which are themselves strongly helical.
These knots exist
where the loops of the trefoil cross and tend to block the evolution of the trefoil. 
However, by forming negative helicity vortex sheets like those in figures 
\ref{fig:tr6T24oh} and \ref{fig:T30knot}, a mechanism exists for circumventing those 
helical patches.  For example, the formation of $h<0$ vortex sheets is balanced by 
simultaneous increases of $h>0$ on the trefoil's cores in \cite{KerrPRF2023}, 
allowing for increases in the vorticity along those cores without violating the 
pre-reconnection, nearly Euler, conservation of helicity.

The new work with initial orthogonal vortices shows that each vortex contributes its own 
sheet, to create a pair. With the compression pushing the vortices together again being 
balanced by viscous diffusion. This results in the inverse slope (a length) on 
the boundaries of each sheet going as $\delta(\nu)\sim\nu^{1/2}$, balanced by expansion 
and stretching in the two perpendicular directions that goes as $\nu^{-1/4}$. 

Does the influence of 
$\nu^{1/4}$ scaling and formation of double vortex sheets extend further in time? It does
in the sense that double vortex sheets are required for the post-reconnection enstrophy 
growth to accelerate sufficiently to get finite-time energy dissipation, 
$\Delta E_\epsilon$ \eqref{eq:dissanom} and is
consistent with how the wrapped structure at the bottom of figure 
\ref{fig:T30knot} at $t=30$ has formed from the two vortex sheets at $t=24$ in figure 
\ref{fig:tr6T24oh}.  The spiral vortex model of \cite{Lundgren1982} assumed only a single 
vortex sheet spiraling around a central vortex, which is distinctly different than the 
paired sheets of both 
this trefoil and orthogonal vortices. 

To accomplish any of these further objectives, new calculations will be required. Starting with
extending the Reynolds number range of the $m>1$ $\Omega_{{\cal }m}(t)$ convergence in
figures \ref{fig:Om2Om4} to the $\nu=1.56\times10^{-5}$ and $\nu=7.8\times10^{-6}$ cases,
which should be re-run in $(8\pi)^3$ domains on $4096^3$, or equivalent, meshes. 
This should provide for both adequate resolution on the small scales and sufficient padding on 
the large scale to control interactions with the periodic boundaries.
\bGree{And if structure function exponents
can be identified that are consistent with the experimental observations, it would
be further proof that singularities are not required to generate either
finite dissipation or Kolmogorov scaling.}

\bGree{
A following question might then be, how can Kolmogorov spectral scaling develop out of the
vortex sheets?
Could some version of the spiral vortex model of \cite{Lundgren1982} provide insight? 
The graphics in figure \ref{fig:T42knot} would support the relevance to some type spiral model,.
which could be the topic of another paper. }

\section*{Acknowledgements} I would like to thank the Isaac Newton Institute 
for Mathematical Sciences for support and hospitality during the programme
{\it Mathematical Aspects of Fluid Turbulence: where do we stand?} in 2022, when work 
on this paper was undertaken and supported by grant number EP/R014604/1. Further 
thanks to E. Titi (Cambridge) who suggested using domain rescaling and J.C. Robinson
(Warwick) who shared his version of the earlier nonlinear
inequality \citep{Constantin1986}. 

\newpage
\appendix
\section{Domain rescaling \label{sec:rescaling}}

Can the empirical evidence for dependence of $\nu_c(\ell)$ \eqref{eq:nuc} be understood 
analytically? These appendices show that how related analytic critical viscosities $\nu_s$, 
originally estimated 
using $(2\pi)^3$ analysis \citep{Constantin1986}, decrease as $\ell$ increases through rescaling of
that analysis to larger $(2\ell\pi)^3$, $\ell>1$,  domains. Which in turn allows the formation of
of {\it dissipation anomalies} as $\ell$ increases.

In the past it has been
claimed that this can be done by supposing extensions into whole space, infinite $\bbR^3$, 
using the ``usual methods''. One version of that approach is to create new outer, 
whole space length scales out of ratios of higher-order Sobolev norms. An 
approach that requires additional estimates that might also depend upon the 
viscosity.

The approach here is incremental, showing how the $\nu_c(\ell)$ change as $\ell$ increases.
The description will be more physically-based than mathematically-based as in
\cite{JCR2021}.

Appendices \ref{sec:squeeze}-\ref{sec:releasenu} show how one can apply scale invariance to
avoid introducing new length scales when extending that analysis 
to larger $V_\ell=(2\ell\pi)^3$ domains. First, one rescales the $V_\ell$ inner products and norms 
into a $(2\pi)^3$ domain; then one can calculate the critical viscosities $\nu_{s_{2\pi}}$ as 
before; and finish by converting the $\nu_{s_{2\pi}}$ into $\nu_s$ in the original $V_\ell$ 
domain. The primary tools for this analysis are the $\dot{H}^s_\ell=\snm{u_\ell}{\dot{s}}{2}$
norms \eqref{eq:uHs} defined in section \ref{sec:inner}.

Before the rescalings are given, appendix \ref{sec:rescalevarb} presents the domain dependent rescaling
of the variables, properties and norms that will be used throughout.  The first rescaling step in 
appendix \ref{sec:squeeze} then applies those rescalings to squeezing the 
$\snm{u_\ell}{\dot{s}+1}{}$ the Navier-Stokes solutions in large $V_{\ell}=(2\ell\pi)^3$ domains
into a $(2\pi)^3$ domain, followed by application of the resulting 
$\snm{u_\lambda}{\dot{s}+1}{}$ $(\lambda =\ell$) to 
approximations of the $H^s_{2\pi}$ Sobolev mathematics developed for 
$(2\pi)^3$ domains. 


Next, appendix \ref{sec:winequ} determines difference inequalities for the $w=v-u$
equations \eqref{eq:weqn}, with appendix \ref{sec:solwinequ} providing an alternative way to 
solve the cubic nonlinearity in those inequalities with integrating factors. 
This is followed by the determination of the critical viscosities 
$\nu_{s_{2\pi}}$ in appendix \ref{sec:usewinequ} using the $H^{s_{2\pi}}$ norms. 
The sequence of steps in appendices \ref{sec:winequ} to \ref{sec:usewinequ} 
follows Chapter 9 of \cite{RRS2016} and is analogous to what was 
previously found for a $(2\pi)^3$ domain \citep{Constantin1986}. 

Finally, those $\nu_{s_{2\pi}}$ are rescaled back to 
the original $(2\ell\pi)^3$ domain in appendix \ref{sec:releasenu},
resulting in decreasing $\nu_s\sim \exp\left(-\int_0^t\lambda^{s+1}
\snm{u_\ell(\tau)}{\dot{s}+1}{}d\tau\right) \lambda^3$ as $\lambda=\ell$ increases, 
with the $\snm{u_\ell(\tau)}{\dot{s}+1}{}$ coming from the original domain 
and the exponential factor dominating over the algebraic $\lambda^3$ factor.  

This result can explain why the small $\nu$ trefoil results with 
convergent $\sqrt{\nu}Z(t)$, and then convergent $\epsilon=\nu Z(t)$, are
allowed by the mathematics. And if extended further by taking $\ell\to\infty$, 
one might be able identify limiting behaviour in $\bbR^3$, known as whole space.

\subsection{Rescaling of variables and norms. \label{sec:rescalevarb}}

For a periodic domain of size $V_\ell=(2\ell\pi)^3=L^3$, $L=2\pi\ell$, \bpurp{a rescaling 
parameter $\lambda=\ell$ is defined that rescales these variables} into a $L_0=2\pi$, $L_0^3=(2\pi)^3$ domain:
\EQL{eq:Ulambda} L_\lambda=L/\lambda=2\pi,\quad U_\lambda=U/\lambda,\quad 
u_\lambda=u/\lambda  \quad\text{and}\quad \nu_\lambda=\nu/\lambda^2 \EN
such that the Reynolds numbers and inverse time/vorticity scales invariant under $\lambda$:
\EQL{eq:Rlambda} Re_\lambda=U_\lambda L_\lambda/\nu_\lambda =Re\,,\quad
\omega_\lambda=\nabla_\lambda\times u_\lambda=(\lambda/\lambda)\omega=\omega\,.
\EN
The inverse lengths and derivatives when mapped into a $L_\lambda$ domain 
are multiplied by $\lambda$ (left), with 
their effect upon $u$ (right) as follows:
\EQL{eq:nablalambda}\begin{matrix} \medskip \p_\lambda=\lambda\p\,,\quad 
\nabla_\lambda=\lambda\nabla,\qquad\nabla^s u\rightarrow \nabla^s_\lambda 
u_\lambda=\lambda^{s-1}\nabla^s u=\lambda^{s-1}u_{,s} 
\end{matrix} \EN
\noindent such that for $\snm{u}{\dot{s}}{}$ in the $V_\ell$ domain 
(left), in the $V_1$ (right) domain one gets
\vspace{-3mm}
\EQL{eq:Wp1} \snm{u_{,s}}{0}{}=\snm{u}{\dot{s}}{}=\left(\int_{{\cal V}_\ell}(\nabla^s{u})^2 d^3x\right)^{1/2} \Rightarrow
\snm{u_\lambda}{\dot{s}}{}=
\left(\int_{{\cal V}_1}
\lambda^{2s-2}(\nabla^s{u_\lambda})^2 d^3x\right)^{1/2} 
=\lambda^{s-1}\snm{u}{\dot{s}}{}
\EN

\subsection{Squeezing, then rescaling, $H^s$ norms \label{sec:squeeze}} 
Now extend this scaling to vorticity-related properties and Sobolev norms.
\EEM\item For the circulation:
\EQL{eq:circlamb} \Gamma_\lambda\sim U_\lambda L_\lambda=\Gamma/\lambda^2\quad{\rm and}
\quad a_\lambda=a/\lambda,.\qquad\qquad\qquad 
\qquad\qquad\qquad\qquad \EN
\item The nonlinear timescale is invariant:
\EQL{eq:TNLlamb} T_{NL}=T_\lambda=a_\lambda^2/\Gamma_\lambda=
(a^2/\lambda^2)/(\Gamma/\lambda^2) =T_{NL}\,.\EN
\item The volume-integrated enstrophy $Z=V\snm{u}{1}{2}\sim\Gamma^2/L$ \bpurp{such that}
\EQL{eq:enstr} Z_\lambda\sim(\Gamma^2/L)=Z/\lambda^3\EN
\item The volume-integrated energy $E=\half V\snm{u}{0}{2}\sim\Gamma^2 L$ so \bpurp{that}
\EQL{eq:estr} E_\lambda\sim\Gamma_2 L/(\lambda^2\lambda^2\lambda)=E/\lambda^5 \EN
\item Finally, the viscous timescale is unchanged:
\EQL{eq:Tnulamb} 
t_\nu=\nu_\lambda\Lap_\lambda u/u\sim\nu/\lambda^2(\lambda^2\Lap u/\lambda)/(u/\lambda)\sim\nu\Lap u\EN
\EEN
{\it Classical result} Does this rescaling affect the H\"older and 
Cauchy-Schwarz inequalities that lead to this time inequality \cite{DoeringARFM2009}?
(see also \cite{JCR2020})
\EQL{eq:unique} \ddt\snm{\nabla u}{0}{2}\leq\frac{c'}{\nu^3}\snm{\nabla u}{0}{6}\,. \EN
$c'$ is constant that is independent of the periodic domain size. From this one gets
for $u_0=u(0)$ 
\EQL{eq:uniqueNL} \snm{\nabla u(\cdot,t)}{0}{2}\leq \dfrac{\snm{\nabla u_0}{}{2}}
{\sqrt{1-2c'\snm{\nabla u_0}{}{4}t/\nu^3}} \EN
Using the denominator of this inequality, the times over which the solutions are ensured 
to be smooth and unique are
\EQL{eq:tb} 0<t<t_b=\frac{\nu^3}{2c'\snm{\nabla u_0}{}{4}} \EN
Because the rescaled viscosity is $\nu^3_\lambda=\nu/\lambda^6$ and enstrophy squared is
$Z_\lambda^2=\snm{\nabla u}{}{4}\sim Z/\lambda^6$, after rescaling 
$t_{b\lambda}=t_b$. \bpurp{Such that} this \bpurp{time scale} \eqref{eq:Tnulamb} is not 
affected by the rescaling \bpurp{with $\lambda$}. 

\subsection{Difference inequalities for $w$ with Euler $u$. \label{sec:winequ}}

For $\ell=1$, by using higher-$s$-order Sobolev time nonlinear inequalities one can find
the $(2\pi)^3$ domain, analytic critical viscosities $\nu_s$ that mark the threshold 
for when $\nu\to0$ Navier-Stokes solutions can be bounded by $\nu\equiv0$ Euler solutions.
Higher-$s$-order Sobolev time nonlinear inequalities are used because they allow us to go beyond 
what the $s=0$ $(d/dt)\half\snm{u}{0}{2}$ equation \eqref{eq:norm} can tell us. The method used is
an alternative to \cite{Constantin1986} outlined by \cite{RRS2016}.
\footnote{The approach in A.5 and A.6 closely follows Chapter 9 of \citep{RRS2016}, with
Lemma 1 of the robustness proof being a version by the solution to problem 9.2.}

The relevant Sobolev time inequalities are higher-order versions of the $s$-order inner products 
between the velocity difference $\bw=\bv-\bu$, higher-order time derivatives $\nabla^s w_t$ and
its equation \eqref{eq:weqn}. 
\EQL{eq:wtold6} \half\ddt\snm{w}{\dot{s}}{2} + \nu\snm{w}{\dot{s}+1}{2}\leq 
\nu\snm{u}{\dot{s}+1}{}\snm{w}{\dot{s}+1}{} +d_s\snm{w}{\dot{s}}{3}
+d_s\snm{u}{\dot{s}}{}\snm{w}{\dot{s}}{2} +c_s\snm{u}{\dot{s}+1}{}\snm{w}{\dot{s}}{2} \EN
The resulting inequality equations have several cubic nonlinearities and two sets of constants.
For order $s$ terms, $c_s$, and for nonlinear order $s+1$ terms, $d_s$.
\EQL{eq:csds}c_s=2^sC_s\quad\text{and}\quad d_s=s2^s C_{s-1}
\quad\text{with}\quad C_s\sim L^{-3/2+s}.  \EN
The goal is to reduce \eqref{eq:wtold6} to an inequality with a single cubic nonlinearity.
Then solve it to show if there are bounds upon Navier-Stokes solutions.

To get \eqref{eq:wtold6}, gradients need to be added to the $B_{NL}(\cdot,\cdot)$ inner products 
in the $w$ equation \eqref{eq:weqn}. These become order-$s$ inner products 
$\langle{uv}\rangle_{\dot{s}}^2$ \eqref{eq:inner}, 
from which $w$-based energy inequalities can be created. For example
\EQL{eq:Buw} \snm{B_{NL}(u,w)}{\dot{s}}{}\leq 
c_s\snm{u}{\dot{s}}{}\snm{w}{\dot{s}+1}{}\,.\EN
To find the analytic $\nu_s$, some reductions are needed.
The next step is make estimates of the higher-$s$, inner 
products of higher-order versions of the three $B_{NL}(\cdot,\cdot)$ nonlinear terms of the $\bw$ 
equation \eqref{eq:weqn}: 
$(\bu\cdot\nabla)\bw$, $(\bw\cdot\nabla)\bu$, $(\bw\cdot\nabla)\bw$ after contraction with 
$\hat{\bw}_{\bk}$ as in \eqref{eq:norm}. This contraction converts these quadratic nonlinearities 
into cubic nonlinearities.

To reduce the multiple cubic terms into a single cubic nonlinearity requires either raising or 
lowering the order of the components terms, a process that through the $c_s$ and $d_s$
\eqref{eq:csds} effectively introduces a length scale of $2\pi$ into the nonlinear inequality 
analysis. If the $s$-order can be retained then $c_s$ is used and only $s>3/2$ analysis
is required.  Starting with how this two $s$-order inner 
products is absorbed between a test function $\bw$ and $B_{NL}(\cdot,\cdot)$. 
\EQL{eq:Bwu} |(B_{NL}(w,u),w)_{\dot{s}}|\leq c_s\snm{u}{\dot{s}+1}{}\snm{w}{\dot{s}}{2}\,.  \EN

However, to handle the nonlinear time inequality needed to find the $\nu_s$ in 
$H_{\dot{s}}$, $d_s$ and $s\geq5/2$ is required. 
A rigorous argument from \cite{ConstantinFoias1988} shows how order
reduction of the inequality analysis can be done when higher-order $s\!>\!5/2$ is used. 
Specifically the order $s\!+\!1$ terms can be reduced to order $s$ by using the constant 
$d_s$ \eqref{eq:csds}, so that the $\snm{\cdot}{\dot{s}+1}{}$ terms in \eqref{eq:Bwu} are 
converted into following $\snm{\cdot}{\dot{s}}{}$. 
\EQL{eq:Bww} |(B_{NL}(w,w),w)_{H^{\dot{s}}}|\leq d_s\snm{w}{\dot{s}}{3} \quad\text{and}\quad
|(B_{NL}(u,w),w)_{H^{\dot{s}}}|\leq d_s\snm{u}{\dot{s}}{}\snm{w}{\dot{s}}{2}\,. \EN
The first term converts the $B_{NL}(w,w)$ term into $\snm{w}{\dot{s}}{3}$, a cubic 
nonlinearity, and the second allows the $B_{NL}(u,w)$ term to be paired with 
$B_{NL}(w,u)$ term to form an integrating factor. 

The next steps reduce the two viscous terms into a single term on the 
right-hand side using Young's inequality
$ ab\leq\half\left(\epsilon a^2+\frac{b^2}{\epsilon} \right)$
with $\epsilon=1/2$ to combine the first term on the right 
with the $\nu\snm{w}{s+1}{2}$ on the left. Then the last two terms are combined by
using Cauchy-Schwarz upon the left one, that is 
$d_s\snm{u}{\dot{s}}{}\leq L c_s\snm{u}{\dot{s}+1}{}$ to get 
\EQL{eq:wtold7} \half\underbrace{\ddt\snm{w}{\dot{s}}{2}}_{L^{5-2s}T^{-3}}\leq
\underbrace{\frac{\nu}{4}\snm{u}{\dot{s}+1}{2}}_{(L^2T^{-1})(L^{3-2s}T^{-2})}
+\underbrace{d_s\snm{w}{\dot{s}}{3}}_{(L^{-5/2+s})(L^{15/2}T^{-3})L^{-3s}}
+\underbrace{2c_s}_{M_s\sim L^{-3/2+s}}
\underbrace{\snm{u}{\dot{s}+1}{}\snm{w}{\dot{s}}{2}}_{(L^{15/2}T^{-3})L^{-3s-1}}\,. \EN
Underbraces are included to demonstrate the consistent order of each of 
its terms after the conversion of the $\snm{\cdot}{\dot{s}+1}{}$
\eqref{eq:Bwu} into $\snm{\cdot}{\dot{s}}{}$ in \eqref{eq:Bww}.
\footnote{From a non-rigorous perspective this effectively adds an inverse length that comes 
from the dependence of $C_s$ \eqref{eq:csds} upon $L^{-1}$.}
\noindent

\subsection{Solving the difference inequalities for $w$ with Euler $u$. \label{sec:solwinequ}}
The last term on the right in \eqref{eq:wtold7} can be used to create
these integrating factors: 
\EQL{eq:intfE} E_s(t)=\exp\left(\int_0^t M_s\snm{u(\tau)}{\dot{s}+1}{} d\tau\right),\quad
M_s=d_sL+c_s=s2^s C_{s-1}L+2^s C_s\,,\EN 
which with $C_s=L^{-3/2+s}$ becomes $M_s\sim s2^s L^{-3/2+s}$. 
This allows \eqref{eq:wtold7} to be rewritten as
\EQL{eq:Ewsold8} \ddt\left(E_s^{-2}(t)\snm{w(t)}{\dot{s}}{2}\right)\leq 
\frac{\nu}{2}\snm{u}{\dot{s}+1}{2}
E_s^{-2}(t)+2d_s E_s(t)\snm{w}{\dot{s}}{3} E_s^{-3}(t)\,. \EN
Then by defining
\EQL{eq:Ewold9} X(t)=E_s^{-2}(t)\snm{w(t)}{\dot{s}}{2},\quad
\alpha(t)=2d_s E_s(t)\quad\text{and}\quad f(t)=\snm{u}{\dot{s}+1}{2} E_s^{-2}(t)\,,\EN
and using the following Lemma, including the restrictions of 
(\ref{eq:Xcond},\ref{eq:Xdencond},\ref{eq:Xtleq}), 
one obtains an improved bound using $X(t)$. 

\paragraph{Lemma 1.} {\it Suppose that $X\geq0$ is continuous and 
\EQL{eq:Xleq}  \dot{X}\leq \alpha(t)X^{3/2} + (\nu/2)f(t), \quad X(0)=0, \EN
for some $\alpha(\cdot)>0$ and $f$ positive and integrable on $[0,T]$. Then 
for $X(t)<\infty$ assume
\EQL{eq:Xcond} \nu\left(\int_0^t\alpha(s)ds\right)\left(\int_0^t f(s)d\right)^{1/2}<1; \EN
and by applying rules for integration by parts to the right-hand-side of \eqref{eq:Xleq} one gets
\EQL{eq:X9} X(t)\leq\frac{(\nu/2)\int_0^t f(\tau)d\tau}{\left[1-(2\nu)^{1/2}\left(\int_0^t\alpha(\tau)d\tau\right)
\left(\int_0^t f(\tau)d\tau\right)^{1/2}\right]^2}\,, \EN
while 
\EQL{eq:Xdencond} 8\nu\left(\int_0^t\alpha(s)ds\right)^2\left(\int_0^t f(s)ds\right)\leq1;
\EN 
and this bound holds:
\EQL{eq:Xtleq} X\leq2\nu\int_0^t f(s)ds\,. \EN }
\\
{\it Proof:} Start with this overestimate: On the time interval $[0,T]$ the 
solutions $X(t)$ of \eqref{eq:Xleq} are bounded above by the solution of 
\EQL{eq:preNLeq} \obdot{Y}=\alpha(t)Y^{3/2},\quad Y(0)=(\nu/2)\int_0^T f(s)ds \EN
Now simply observe that the solution $Y(t)$ of \eqref{eq:preNLeq} satisfies
$$ \frac{1}{Y(0)^{1/2}} - \frac{1}{Y(t)^{1/2}} = 2\int_0^t\alpha(s)ds,$$
and hence 
\EQL{eq:Yt} Y(t)\leq\frac{Y(0)}{\left[1-\dint_0^t\alpha(s)ds\left(2Y(0)^{1/2}\right)
\right]^2}\,.\EN
Finally, by defining $Y(0)$ using \eqref{eq:preNLeq} one gets \eqref{eq:X9}.
\hfill${\pmb\square}$


\subsection{Using the difference inequality solutions \label{sec:usewinequ}}
By re-inserting the definitions for $\alpha(t)$ and $f(t)$ \eqref{eq:Ewold9} into their
time integrals \eqref{eq:Xdencond}, one 
arrives at these conditions for critical viscosities $\nu_s$: 
\footnote{This determination of the $\nu_s$, and the discussion in A.5 and A.6, follow
earlier discussions with J. C. Robinson.}
\EQL{eq:muleq} 8\nu_s\left(\int_0^t 2d_s E_s(\tau)d\tau \right)^2 
\int_0^t\snm{u(\tau)}{\dot{s}+1}{2}E_s^{-2}(\tau)d\tau= 1 \,, \EN
such that all $\nu\leq\nu_s$ Navier-Stokes solutions are 
bounded by the Euler ($u$) solution when 
\EQL{eq:mucrit} \nu\leq \nu_s(t)= \frac{1}{8}\frac{1}{\underbrace{
\left(\dint_0^t 2d_s E_s(\tau)d\tau \right)^2}_{L^{-5+2s}T^2}
\underbrace{
\dint_0^t\snm{u(\tau)}{\dot{s}+1}{2}E_s^{-2}(\tau)d\tau}_{L^{3-2s}T^{-1}}}
\,, \EN
with the dimensions in the underbraces. When multiplied, these terms have the dimension
of an inverse viscosity. Note that for any $\nu$, at very early times when
$\snm{w(t)}{\dot{s}}{}=\snm{v-u}{\dot{s}}{}$ and the $X(t)$ are small,
these relations are irrelevant. 

When critical viscosities $\nu_s$ equivalent to \eqref{eq:mucrit} were first 
derived \citep{Constantin1986}, the goal was to demonstrate the existence of 
this bound. Given our current understanding of the underlying Euler norms, 
can \eqref{eq:mucrit} be used to estimate the $\nu_s$ in terms of time 
integrals of $\alpha(t)$ and $f(t)$?

A first step is to replace the time integrals with approximations at the ends of their 
respective time spans, with the $f(t)$ integral using $E_s^{-2}(t)$ focusing upon small times 
and the integral with $E_s(t)$ focusing upon $d\tau\sim t$.

\noindent$\bullet$ For the integral with $E_s^{-2}$, note that $E_s^{-2}(t=0)=1$, 
so for $t\sim0$ one gets
$$ E_s^{-2}(t)=\exp\left(-2\int_0^t M_s\|u(\tau)\|_{\dot{s}+1}d\tau
\right) \sim{\cal O}(1) $$
and can take $\Delta\tau\sim(2M_s\snm{u_0}{\dot{s}+1}{})^{-1}$ as $E_s^{-2}(t)$ becomes 
exponentially small at large $t$. \\
$\bullet$ For $t\sim0$, and by taking $f(t)$ from \eqref{eq:Ewold9}, one gets
\EQL{eq:Dtau0} \int f(\tau)d\tau\approx\dint_0^t\snm{u(\tau)}{\dot{s}+1}{2}E_s^{-2}(\tau)d\tau
\sim \snm{u(0)}{\dot{s}+1}{2}/(2M_s\snm{u_0}{\dot{s}+1}{}) 
=\snm{u(0)}{\dot{s}+1}{}/(2M_s)\,. \EN 
$\bullet$ For large $t$ in $\alpha(t)$ \eqref{eq:Ewold9} and assuming that for the time span in
question the $|u(\tau)\|_{\dot{s}+1}$ is very large, the focus is upon $\tau\sim t$, with
the largest contributions to
$$ E_s(t)=\exp\left(\int_0^t M_s\|u(\tau)\|_{\dot{s}+1}d\tau
\right) \quad\mbox{coming over the last}\quad \Delta\tau\sim(M_s\snm{u(t)}{\dot{s}+1}{})^{-1}\,.$$ 
$\bullet$ And for $d\tau\sim\Delta\tau$, this results in 
\EQL{eq:Dtaut} \dint_0^t \alpha(\tau)d\tau=\dint_0^t 2d_s E_s(\tau)d\tau \sim 2d_s E_s(t)/(M_s\snm{u(t)}{\dot{s}+1}{})\,. \EN
Putting \eqref{eq:Dtau0} and \eqref{eq:Dtaut} together in \eqref{eq:mucrit},
that is determining $(8(\int_0^t\alpha(t)dt)^2(\int_0^t f(t)dt)^{-1}$, the 
resulting $(2\pi)^3$ domain, $s$-order $\nu_{s_{2\pi}}$ bounds for $\nu$ are 
given by:
\EQL{eq:muapprox}\nu\leq \nu_{s_{2\pi}}(t)\sim 
\left[8\left(2d_s E_s(t)\left(M_s\snm{u(t)}{\dot{s}+1}{}\right)^{-1}\right)^2
\left(\snm{u(0)}{\dot{s}+1}{}\right)\left(2M_s)\right)^{-1}\right]^{-1} \EN
Combining the terms, gives this estimate
\EQL{eq:muapproxt}\nu_{s_{2\pi}}(t)\lesssim
\frac{ M_s^3\snm{u(t)}{\dot{s}+1}{2}}
{16d_s^2 \snm{u(0)}{\dot{s}+1}{} }\EN

Finally, for $\nu\leq\nu_{s_{2\pi}}$ the $\snm{w(t)}{\dot{s}}{2}$ differences between 
the Navier-Stokes and Euler solutions are governed by:
\EQL{eq:wbnd} \snm{w(t)}{\dot{s}}{2}\leq
2\nu\int_0^t\snm{u(r)}{\dot{s}+1}{2}
\exp\left\{2\dint_r^t M_s\snm{u(\tau)}{\dot{s}+1}{} d\tau\right\}dr\,. \EN
Meaning that the $\nu\leq\nu_{s_{2\pi}}$ Navier-Stokes solutions are bounded by 
these integrals of the Euler solutions. So unless those Euler solutions have 
finite-time singularities, and \cite{MengYang2024} suggests they do not, those Euler integrals 
will always be bounded, which 
in turn bounds the $\nu\leq\nu_{s_{2\pi}}$ Navier-Stokes solutions and in particular 
the dissipation rate $\nu Z(t)\to0$.  

$\pmb{\therefore}$ Showing that Navier-Stokes solutions in 
$(2\pi)^3$ domains cannot develop finite energy dissipation in a finite 
time \eqref{eq:dissanom} as $\nu\to0$. 

The next appendix extends this result to arbitrarity large $(2\ell\pi)^3$ 
domains, with the $\nu_s\to0$ as $\ell$ grows. Which would allow a 
dissipation anomaly \eqref{eq:dissanom}.
\subsection{Release squeezing \label{sec:releasenu}}

To estimate the $\nu_s$ critical viscosities for a compact initial condition 
in larger domains $(2\ell\pi)^3$, the first step is to squeeze that compact 
state by $\lambda=\ell$ into a $(2\pi)^3$ domain. This also  means rescaling 
the velocities and derivatives as in \eqref{eq:Ulambda} and the higher-$s$ 
norms as in \eqref{eq:Wp1} giving
\EEM\item 
$\snm{u_\lambda(\tau)}{s+1}{}=\lambda^s\snm{u_\ell(\tau)}{s+1}{}$ 
\eqref{eq:nablalambda} for both $\tau=0$ and $\tau=t$. 
\item Next, create $(2\pi)^3$ integrating 
factors $E_{s,\lambda}(t)$ by putting the $\snm{u_\lambda(t)}{s+1}{}$ in
$E_s$ \eqref{eq:intfE}.  
\item[] Note that because this operation is in 
a $L_0^3=(2\pi)^3$ domain, the $M_s$ terms in $E_{s,\lambda}(t)$ 
do not need to be rescaled.  
\item Then apply $\snm{u_\lambda}{s+1}{}$ and $E_{s,\lambda}$ 
to \eqref{eq:muapprox} to get $\nu_{s_{2\pi}}$. 
\item And use the inverse of $\nu_\lambda=\nu/\lambda^2$ \eqref{eq:Ulambda},
with $\nu_\lambda=\nu_{s_{2\pi}}$ and $\nu=\nu_s$: giving $\nu_s=\lambda^2\nu_{s_{2\pi}}$. 
\item \bpurp{Leading to the following} $\nu_s$ in the original domain. 
\EEN

\EQL{eq:mucritlamb} \nu_s(t)\sim 
\exp\left(-\int_0^t\lambda^{s+1}\snm{u_\ell(\tau)}{\dot{s}+1}{}d\tau\right)
\lambda^3\frac{M_s^3\snm{u_\ell(t)}{\dot{s}+1}{2}}
{16d_s^2\snm{u_\ell(0)}{\dot{s}+1}{}} \EN
with the exponentially-decreasing factor dominating over the algebraic 
$\lambda^3$ term.
This decrease of the $\nu_s$ as $\ell$ increases is 
more than fast enough to support the relevance of the variable domain 
calculations in figure \ref{fig:iQsnuZ}.


\end{document}

%% file: kerrcommands.tex
\newcommand{\bminil}[1]{\begin{minipage}[l]{#1 \textwidth}}
\newcommand{\bminir}[1]{\begin{minipage}[r]{#1 \textwidth}}
\newcommand{\bminic}[1]{\begin{minipage}[c]{#1 \textwidth}}
\newcommand{\emini}{\end{minipage}}
\newcommand{\EQL}{\begin{equation}\label}
\newcommand{\EQ}{\begin{equation}}
\newcommand{\EN}{\end{equation}}
\newcommand{\BFG}{\begin{figure}}
\newcommand{\EFG}{\end{figure}}
\newcommand{\ITM}{\begin{itemize}}
\newcommand{\ITN}{\end{itemize}}
\newcommand{\EEM}{\begin{enumerate}}
\newcommand{\EEN}{\end{enumerate}}
\newcommand{\BEA}{\[\begin{array}}
\newcommand{\EEA}{\end{array}\]}
\newcommand{\EQAL}{\begin{eqnarray}\label}
\newcommand{\EQA}{\begin{eqnarray}}
\newcommand{\ENA}{\end{eqnarray}}

\newcommand{\dint}{\displaystyle \int}

\newcommand{\btriangle}{\mbox{\boldmath$\triangle$}}

\newcommand{\obdot}[1]{\overset{\mbox{\boldmath$.$}}{#1}}
\newcommand{\bI}{\mbox{\boldmath$I$}}

\newcommand{\bP}{\mbox{\boldmath$P$}}

\newcommand{\bS}{\mbox{\boldmath$S$}}

\newcommand{\bk}{\mbox{\boldmath$k$}}

\newcommand{\bn}{\mbox{\boldmath$n$}}

\newcommand{\br}{\mbox{\boldmath$r$}}
\newcommand{\bs}{\mbox{\boldmath$s$}}

\newcommand{\bu}{\mbox{\boldmath$u$}}
\newcommand{\bv}{\mbox{\boldmath$v$}}
\newcommand{\bw}{\mbox{\boldmath$w$}}
\newcommand{\bx}{\mbox{\boldmath$x$}}

\newcommand{\bomega}{\mbox{\boldmath$\omega$}}

\newcommand{\bxi}{\mbox{\boldmath$\xi$}}

\newcommand{\p}{\partial}

\newcommand{\ppto}[1]{\frac{\partial #1}{\partial t}}

\newcommand{\half}{\mbox{$\frac{1}{2}$}}

\newcommand{\ddt}{\frac{d}{dt}}

\renewcommand{\theenumi}{\alph{enumi}}

\newcommand{\bbR}{\mathbb{R}}

\newcommand{\bbZ}{\mathbb{Z}}
